\documentclass[aps,prd,amssymb,eqsecnum,showpacs,nofootinbib,floatfix,twocolumn,a4paper]{revtex4-1}

\usepackage{amsmath}
\usepackage{graphicx}

\usepackage{color}

\allowdisplaybreaks

\newcommand{\be}{\begin{equation}}
\newcommand{\ee}{\end{equation}}

\newcommand{\Pf}{{\mathrm{Pf}}}

\newcommand{\md}{\mathrm{d}}
\newcommand{\me}{\mathrm{e}}

\newcommand{\vecr}{\mathbf{r}}

\newcommand{\vecp}{\mathbf{p}}
\newcommand{\vecpa}{\mathbf{p}_a}

\newcommand{\vecx}{\mathbf{x}}
\newcommand{\vecxa}{\mathbf{x}_a}

\newcommand{\vecre}{\mathbf{r}'}
\newcommand{\vecpe}{\mathbf{p}'}

\newcommand{\nepe}{\mathbf{n}'\cdot\mathbf{p}'}

\newcommand{\Q}{\hat Q}

\newcommand{\gE}{{\gamma_\textrm{E}}}

\newcommand{\npi}{(\mathbf{n}_{12}\cdot\mathbf{p}_1)}
\newcommand{\npii}{({\bf n}_{12}\cdot{\bf p}_2)}

\newcommand{\hTT}[2]{h^\mathrm{TT}_{#1#2}}
\newcommand{\hTTdot}[2]{\dot{h}^\mathrm{TT}_{#1#2}}
\newcommand{\pit}[2]{\tilde{\pi}^{#1#2}}

\begin{document}

\title{Conservative dynamics of two-body systems at the \\
 fourth post-Newtonian approximation of general relativity}

\author{Thibault Damour}
\email{damour@ihes.fr}
\affiliation{Institut des Hautes Etudes Scientifiques, 35 route de Chartres, 91440 Bures-sur-Yvette, France}

\author{Piotr Jaranowski}
\email{p.jaranowski@uwb.edu.pl}
\affiliation{Faculty of Physics,
University of Bia{\l}ystok,
Cio{\l}kowskiego 1L, 15--245 Bia{\l}ystok, Poland}

\author{Gerhard Sch\"afer}
\email{gos@tpi.uni-jena.de}
\affiliation{Theoretisch-Physikalisches Institut,
Friedrich-Schiller-Universit\"at Jena,
Max-Wien-Pl.\ 1, 07743 Jena, Germany}

\date{8 March, 2016}

\begin{abstract}
The  fourth post-Newtonian (4PN) two-body dynamics has been recently tackled
by several different approaches: effective field theory, Arnowitt-Deser-Misner Hamiltonian,
action-angle-Delaunay averaging, effective-one-body, gravitational self-force, first law of dynamics,
and Fokker action. We review the achievements of these approaches
and discuss the complementarity of their results.
Our main conclusions are: (i) the results of the first complete derivation of the 4PN dynamics 
[T.~Damour, P.\ Jaranowski, and G.\ Sch\"afer, Phys.\ Rev.\ D {\bf 89}, 064058 (2014)]
have been, {\it piecewise}, fully confirmed by several subsequent works;
(ii) the results of the Delaunay-averaging technique
[T.\ Damour, P.\ Jaranowski, and G.\ Sch\"afer, Phys.\ Rev.\ D {\bf 91}, 084024 (2015)]
have been confirmed by several independent works;
and (iii) several claims in a recent harmonic-coordinates Fokker-action computation
[L.~Bernard \textit{et al.\@}, arXiv:1512.02876v2 [gr-qc]] are incorrect,
but can be corrected by the addition of a couple of {\it ambiguity parameters} linked to subtleties 
in the regularization of infrared and ultraviolet divergences.
\end{abstract}

\maketitle

\section{Introduction}

The general-relativistic two-body problem has acquired a renewed importance
in view of the impending prospect of detecting the gravitational wave signals emitted by inspiraling and coalescing compact binaries.
After the successful completion, 15 years ago, of the derivation of the third post-Newtonian (3PN) two-body dynamics \cite{Damour:2001bu} 
(see Refs.\ \cite{Itoh:2003,Itoh:2004,Blanchet:2004,Foffa:2011} for later rederivations),
a natural challenge was to tackle the (conservative) fourth post-Newtonian (4PN)  dynamics.
Several partial steps in the derivation of the 4PN dynamics have been performed
\cite{Damour:2009sm,Blanchet:2010zd,Damourlogs,Barausse:2011dq,Jaranowski:2012eb,Foffa:2012rn,Jaranowski:2013lca,Bini:2013zaa,Jaranowski:2015lha}\footnote{
See also Ref.\ \cite{Ledvinka:2008} for the $O(G^1)$ interaction, valid to {\it all} PN-orders.}, and culminated in
the recent first full derivation, within the Arnowitt-Deser-Misner (ADM) Hamiltonian formulation of general relativity,  of the 4PN dynamics in Ref.\ \cite{Damour:2014jta}, completed by an action-angle-Delaunay averaging study in Ref.\ \cite{Damour:2015isa}. A remarkable feature of the
conservative 4PN dynamics of Ref.\ \cite{Damour:2014jta} is the presence of a {\it time-symmetric nonlocal-in-time} interaction, which is directly
related to the 4PN-level tail-transported (retarded) interaction first discussed in Ref.\ \cite{Blanchet:1987wq}
(see also Refs.\ \cite{Foffa:2011np,Galley:2015kus} for recent rediscussions).

In a recent preprint \cite{Bernard:2015njp} Bernard \textit{et al.\@} reported the computation of the Fokker action describing the 4PN dynamics  in harmonic coordinates. They made several claims in (version 2 of) their preprint, notably (i) the presence of a {\it unique} ambiguity, of infrared (IR) origin, parametrized by the parameter $\alpha$, and (ii) the need to choose the value $\alpha = \alpha_{\rm B^3FM} = 811/672$ to reproduce the analytically known \cite{Bini:2013zaa} 4PN interaction energy for circular orbits.
Moreover, they stated that their 4PN dynamics disagrees with the one recently derived by us \cite{Damour:2014jta,Damour:2015isa,Jaranowski:2015lha} and that part of the discrepancy comes from our treatment of the nonlocal contribution to the dynamics.

The first aims of the present paper will be:
1) to show  that the claims, denoted as (i) and (ii) above, of Ref.\ \cite{Bernard:2015njp} are incorrect;  
2) to suggest that the discrepancies between their result and ours
is due both to their incorrect evaluation of the conserved energy of the nonlocal 4PN dynamics 
and to the need to complete their result by at least one further contribution parametrized by an additional ambiguity parameter,
denoted $a$ below.
We, however, emphasize that our suggested corrections
represent only two rather minor adjustments among hundreds of terms that agree between 
two very difficult (four-loop level!) independent calculations, using different methods and different gauges.
The other aims of our paper will be:
3) to explain in more detail than in our original paper  \cite{Damour:2015isa} the logical basis and consistency 
of our reduction of the nonlocal 4PN dynamics to a formally local action-angle Hamiltonian;
and 4) to summarize the many independent results
that have confirmed, piecemeal, all the elements of  our 4PN  dynamics \cite{Damour:2014jta,Damour:2015isa}.

Regarding the last item, we wish to emphasize that,
besides the straightforward PN (or, equivalently, effective-field-theory) calculations
of the two-body dynamics, crucial information about the two-body dynamics has been acquired
(as will be detailed below) through combining (in various ways) several other approaches: 
the effective-one-body (EOB) formalism \cite{Buonanno:1998gg,Buonanno:2000ef,Damour:2000we,Damour:2001tu},
the first law of binary mechanics \cite{LeTiec:2011ab,Blanchet:2012at,Tiec:2015cxa},
and gravitational self-force (SF) theory,
especially when combined with the Mano-Suzuki-Takasugi \cite{Mano:1996mf,Mano:1996vt}
hypergeometric-expansion approach to Regge-Wheeler-Zerilli theory.
We shall point out when needed the multi-way consistency checks between these approaches
that have been obtained at the 4PN level.

Our notation for two-body systems will follow the one used in our previous works:
the two masses are denoted $m_1$, $m_2$, and we then denote
$M \equiv m_1 + m_2$, $\mu \equiv m_1 m_2/(m_1+m_2)$, $\nu \equiv \mu/M = m_1 m_2/(m_1+m_2)^2$.
We indicate powers of $G$ and $c$ when it is pedagogically useful,
but we sometimes set $G=c=1$ when it is more convenient.

\section{Extant piecewise confirmations of the 4PN ADM-Delaunay dynamics and limits on possible deviations}

Let us start by recalling the structure of the 4PN results
of Refs.\ \cite{Damour:2014jta,Damour:2015isa}.
First, the Fokker-like reduced\footnote{See Appendix A for a discussion of Fokker-like reduced actions.}
ADM action derived in Ref.\ \cite{Damour:2014jta} has the following form
(using the notation of Ref.\ \cite{Damour:2014jta}):
\begin{align}
\label{SDJS}
S_\textrm{DJS} &= \int \Big[\sum_a p_{ai}\,\md x^i_a
- H^\textrm{loc}_\textrm{DJS}(\vecxa,\vecpa;s)\,\md t\Big]
\nonumber\\&\quad
+ S^\textrm{nonloc ($s$)},
\end{align}
where the {\it nonlocal} piece of the ADM action\footnote{In Ref.\ \cite{Damour:2014jta} the necessity of adding the
nonlocal contribution $S^\textrm{nonloc ($s$)}$ was derived by combining the structure of the IR divergence of the local ADM action 
\cite{Jaranowski:2015lha} with the known existence of a 4PN-level, long-range tail-transported interaction \cite{Blanchet:1987wq}. 
Prompted by an argument in Ref.\ \cite{Bernard:2015njp}, we discuss in Appendix A the (limited) extent
to which $S^\textrm{nonloc ($s$)}$ can be (formally) directly derived from the ADM action.}
is (with $I_{ij}$ denoting the quadrupole moment of the binary system,
and $I_{ij}^{(3)}\equiv\md^3 I_{ij}/\md t^3$)
\be
\label{Snonloc}
S^\textrm{nonloc ($s$)} = \frac{1}{5}\frac{G^2M}{c^8}
\Pf_{2s/c} \iint \frac{\md t\,\md t'}{|t-t'|}I_{ij}^{(3)}(t)I_{ij}^{(3)}(t') \, ,
\ee
(${\rm Pf} \equiv$ Partie finie) and where the {\it local} piece of the ADM action has the structure
\begin{align}
\label{eq4}
H^\textrm{loc}_\textrm{DJS}(\vecxa,\vecpa;s)
&= H^\textrm{loc}_\textrm{DJS}(\vecxa,\vecpa;r_{12})
\nonumber\\[1ex]&\quad
+ F[\vecxa,\vecpa]\ln\frac{r_{12}}{s},
\end{align}
with
\be
\label{eq6}
F[\vecxa,\vecpa] = \frac{2}{5}\frac{G^2M}{c^8}
(I_{ij}^{(3)})^2.
\ee
The length scale $s$ entering the action above\footnote{For clarity,
we do not use here the formulation where $s$ is replaced by $r_{12}=\vert\vecx_1-\vecx_2 \vert$.}
is arbitrary because (as shown in Ref.\ \cite{Damour:2014jta})
the $s$ dependence of the nonlocal action \eqref{Snonloc}
cancels against the $s$ dependence of the local action \eqref{eq4}.

In our second paper, Ref.\ \cite{Damour:2015isa}, we combined different techniques (which will be explained in detail below)
for transforming the (center-of-mass-frame reduction of the) nonlocal ADM action \eqref{SDJS}  into an equivalent ordinary action, say
\begin{align}
\label{SDJS'}
S_{\textrm{DJS}'} &= \int \Big[ p_{i}\,\md q^i
- H_{\textrm{DJS}'}(q, p)\,\md t\Big],
\end{align}
where $H_{\textrm{DJS}'}(q, p)$ is formally given by an expansion in (even) powers of $ q^i p_i $.
[Reference \cite{Damour:2015isa} gave an exact formula for the action-angle version of $S^\textrm{nonloc}$,
and the explicit form of the expansion of $H_{\textrm{DJS}'}(q, p)$ in powers of $ q^i p_i $ through $ (q^i p_i)^6 $.]
The ordinary action \eqref{SDJS'} is equivalent to (the center-of-mass-frame reduction of) Eq.\ \eqref{SDJS}
modulo some shifts of the phase-space coordinates that are implicitly defined
by the Delaunay-like reduction procedure of Ref.\ \cite{Damour:2015isa}.
These shifts do not affect the gauge-invariant observables deduced from either Eq.\ \eqref{SDJS} or Eq.\ \eqref{SDJS'}
which we shall focus on in the following.
Note also in passing that the use of the EOB formalism in Ref.\ \cite{Damour:2015isa}
is essentially a technical convenience, while the essential conceptual step used there is the reduction of a nonlocal action to a local form by
using Delaunay-like averaging techniques (see below). 

The aim of this section is to summarize the current existing confirmations of the correctness of the 4PN actions \eqref{SDJS} or \eqref{SDJS'}.
First, we wish to recall that the full Poincar\'e invariance of \eqref{SDJS} (in a general frame)
was explicitly checked in Ref.\ \cite{Damour:2014jta} (see also Ref.\ \cite{Jaranowski:2015lha}).
This is a highly nontrivial check because the ADM derivation of the action (contrary to the harmonic-coordinates one)
is far from being manifestly Lorentz invariant. 

To organize the other confirmations of Eq.\ \eqref{SDJS} or Eq.\ \eqref{SDJS'}, let us note that, in the center-of-mass frame,
and when using suitably scaled\footnote{Beware that we will often oscillate between using scaled or unscaled dynamical variables.}
dynamical variables [e.g.\ $r= r^{\rm phys}/(GM)$, $\vecp = \vecp_1/\mu=- \vecp_2/\mu$,
$S=S^{\rm phys}/(GM\mu)$, $\widehat H=(H- Mc^2)/\mu$]
the rescaled Hamiltonian  $\widehat H$ has a polynomial structure in the symmetric mass ratio $\nu$
\be
\label{nuexpandedH}
\widehat H = \widehat H_0 + \nu \widehat H_1+ \nu^2 \widehat H_2 + \nu^3 \widehat H_3 + \nu^4 \widehat H_4,
\ee
where $\widehat H_0 = \frac12 \vecp^2 -1/r + c^{-2}( \cdots) + \cdots + O(c^{-10})$ 
describes the 4PN dynamics of a test particle in the field generated by the mass $M=m_1+m_2$.

The contribution $ \nu \widehat H_1$ to the 4PN dynamics describes the 4PN approximation to the first-order self-force (1SF) dynamical
effects. Over the recent years, many works have been devoted to both the numerical and the analytical computation of
1SF dynamical effects. To compare these results to the predictions following from our 4PN dynamics \eqref{SDJS} or \eqref{SDJS'} we need
{\it bridges} between PN results and SF results. Two such PN-SF bridges have been particularly useful over the last years: the EOB formalism
\cite{Buonanno:1998gg,Buonanno:2000ef,Damour:2000we,Damour:2001tu}, and 
 the first law of binary mechanics \cite{LeTiec:2011ab,Blanchet:2012at,Tiec:2015cxa}. 
 
The first example of a PN-EOB-SF bridge was the derivation of the functional relation
between the periastron precession of small-eccentricity orbits
and the radial potentials entering the EOB Hamiltonian \cite{Damour:2009sm}.
At the 1SF level, this relation [see Eqs.\ (5.21)--(5.25) in \cite{Damour:2009sm}]
yields the precession function $\rho(u)$ as a linear combination of $a(u)$, $a'(u)$, $a''(u)$ and $\bar d(u)$,
where $a(u)$ is the 1SF correction to the main EOB radial potential $A(u;\nu)= 1- 2u + \nu a(u) + O(\nu^2)$
[which generalizes $A^{\rm Schwarzschild} (u) = 1-2 GM/c^2 r \equiv 1-2u$], 
and where $\bar d(u)$ is the 1SF correction to the second EOB radial potential
$\bar D(u) \equiv ( A(u) B(u))^{-1} = 1 + \nu \bar d(u)  + O(\nu^2)$.
Let us apply this relation to the 4PN-level values of the EOB potentials
parametrizing the 4PN dynamics \eqref{SDJS'}, as derived in Ref.\ \cite{Damour:2015isa}, namely (adding the third EOB potential $Q\equiv\mu^2\hat{Q}$)
\begin{widetext}
\begin{subequations}
\label{ADQ}
\begin{align}
A(u) &= 1 - 2 u + 2 \nu \,u^3 + \left(\frac{94}{3}-\frac{41\pi^2}{32}\right) \nu \,u^4
\nonumber\\[1ex]&\quad + \bigglb(
\left( \frac{2275\pi^2}{512} - \frac{4237}{60} + \frac{128}{5}\gE + \frac{256}{5}\ln2 \right) \nu
+ \left( \frac{41\pi^2}{32} - \frac{221}{6} \right)\nu^2
+ \frac{64}{5}\nu\,\ln u \biggrb) u^5,
\\[2ex]
\bar{D}(u) &= 1 + 6\nu\,u^2 + \left(52\nu - 6\nu^2\right) u^3
\nonumber\\[1ex]&\quad + \bigglb(
\left( -\frac{533}{45} - \frac{23761\pi^2}{1536} + \frac{1184}{15}\gE - \frac{6496}{15}\ln2 + \frac{2916}{5}\ln3 \right)\nu
+ \left(\frac{123\pi^2}{16} - 260\right)\nu^2 + \frac{592}{15}\nu\ln u \biggrb) u^4,
\\[2ex]
\Q(\vecre,\vecpe) &= \bigglb( 2(4-3\nu )\nu\,u^2
+ \left( \left( -\frac{5308}{15} + \frac{496256}{45}\ln2 - \frac{33048}{5}\ln3 \right) \nu
- 83 \nu^2 + 10 \nu^3 \right) u^3 \biggrb) (\nepe)^4
\nonumber\\[1ex]&\quad
+ \bigglb( \left(-\frac{827}{3} - \frac{2358912}{25}\ln2 + \frac{1399437}{50}\ln3
+ \frac{390625}{18}\ln5 \right)\nu - \frac{27}{5}\nu^2 + 6\nu^3 \biggrb)u^2\,(\nepe)^6 + O[ \nu u  (\nepe)^8].
\end{align}
\end{subequations}
This yields
\begin{align}
\rho(x)=14 x^2 &+ \left(\frac{397}{2}-\frac{123}{16}\pi^2\right) x^3
\nonumber\\[1ex]
&+\left( \frac{58265}{1536}\pi^2 - \frac{215729}{180} +\frac{5024}{15}\gE
+ \frac{1184}{15}\ln2 + \frac{2916}{5}\ln3 
+ \frac{2512}{15}\ln x \right) x^4
+ O( x^5 \ln x).
\end{align}
\end{widetext}
The 4PN-level contribution to the precession function $\rho(x)$  is of the form
$ \rho_{4 \rm PN}(x)= (\rho_4^c + \rho_4^{\ln} \ln x) \,  x^4$, with the rational logarithmic coefficient 
$\rho_4^{\ln} =\frac{2512}{15}$ \cite{Damourlogs},
and the transcendental nonlogarithmic 4PN coefficient
\begin{multline}
\label{rho4djs1}
\rho_4^{c, \rm DJS}= \frac{58265}{1536}\pi^2 - \frac{215729}{180} 
\\[1ex]
+ \frac{5024}{15}\gE
+ \frac{1184}{15}\ln2 + \frac{2916}{5}\ln3 .
\end{multline}
The {\it analytical} values of the 4PN-level functions $A$, $\bar D$, $Q$ have been recently independently confirmed
by two self-force computations (based on the recently derived eccentric-extension of the first law \cite{Tiec:2015cxa});
see Refs.\ \cite{Bini:2015bfb} and \cite{Hopper:2015icj}. 
(Note also that the coefficients of higher powers of $\nepe$ in $Q$ were recently provided:
the powers 8 and 10 in Ref.\ \cite{Hopper:2015icj}; and the powers 12 to 20 in Ref.\ \cite{BiniDamourGeralico2016}.) 

In addition to analytical confirmations, there are also numerical self-force confirmations of the 4PN-level values of $\rho (x)$ and $\rho_4^{c, \rm DJS}$ based on direct dynamical computations of the precession of slightly eccentric orbits.

The first numerical SF determination of the precession function $\rho(u)$ was made in Ref.\ \cite{Barack:2010ny}.
In particular, the latter work confirmed the value $\rho_4^{\ln} =\frac{2512}{15}$
and derived an estimate of the value of the nonlogarithmic coefficient, namely
\be
\label{rho4barack}
\rho_4^{c, \textrm{num\cite{Barack:2010ny}}} = 69^{+ 7}_{- 4}.
\ee 
We were informed  by Maarten van de Meent that he has very recently obtained a much more accurate determination 
of $\rho_4^{c}$ with {\it preliminary} results yielding  \cite{vandeMeent:2016}
\be
\label{rho4meent}
\rho_4^{c, \textrm{num\cite{vandeMeent:2016}}} = 64.640566(2),
\ee
where the number in parentheses indicates a preliminary estimate of the uncertainty on the last digit. 
Note that both numerical estimates of  $\rho_4^{c}$ assume the analytical values of the 4PN and 5PN logarithmic contributions to $\rho(u)$.
[There is no doubt about the (1SF) 4PN and 5PN  logarithmic contributions to the dynamics:
the 4PN ones can be straightforwardly deduced from the tail-related 4PN logarithmic term written in Eq.\ (6.39) of Ref.\ \cite{Blanchet:1987wq}
(equivalent to the $F[\vecxa,\vecpa]\ln\frac{r_{12}}{s}$ logarithmic contribution in Eq.\ \eqref{eq4}),
while the 5PN ones are straightforwardly derivable from the higher-tail results
of Refs.\ \cite{Blanchet:2010zd} and \cite{Damour:2015isa} (Sec.\ IXA).]
Both numerical SF results \eqref{rho4barack} and \eqref{rho4meent} confirm, within their respective error bars, the numerical value
\be
\label{rho4djs2}
\rho_4^{c, \rm DJS} = 
64.64056\, 47571\, 19378\, 19014\, 84255\cdots
\ee
of the result \eqref{rho4djs1} predicted by our 4PN dynamics.
The numerical result \eqref{rho4barack} differs from the analytical value \eqref{rho4djs1} by
\be
\label{drho4barack}
\Delta \rho_4^{c} \equiv \rho_4^{c, \textrm{num\cite{Barack:2010ny}}}
- \rho_4^{c, \rm DJS} = 4^{+ 7}_{- 4},
\ee
while the more recent numerical value \eqref{rho4meent}
differs from the analytical value \eqref{rho4djs1} only by 
\be
\label{drho4meent}
\Delta \rho_4^{c} \equiv \rho_4^{c, \textrm{num\cite{vandeMeent:2016}}}
- \rho_4^{c, \rm DJS}=(1 \pm 2) \times 10^{-6}. 
\ee
What is especially important in such a numerical check is that we are talking here
about a direct dynamical check of the 4PN dynamics derived in Refs.\ \cite{Damour:2014jta} and \cite{Damour:2015isa}.
Indeed, no use is made here of the first-law-of-mechanics bridge to go from SF computations of the Detweiler-Barack-Sago redshift invariant to 4PN dynamical functions.
The SF computations done in Refs.\ \cite{Barack:2010ny} and \cite{vandeMeent:2016}
directly estimated the effect of the nonlocal gravitational self-force $F^{\mu}$
on slightly eccentric orbits to extract the precession function $\rho(u)$ and, then, its PN expansion coefficients.
One has therefore here a direct confirmation of the way Refs.\ \cite{Damour:2014jta,Damour:2015isa} computed
(via well-established EOB results \cite{Damour:2009sm}) the precession effect of the nonlocal 4PN dynamics. 

Let us now summarize the numerical SF confirmations of the other 1SF predictions one can draw from the 4PN results \eqref{ADQ}.
The 4PN-level coefficients of the three EOB potentials $A$, $\bar D$ and $Q$
[the latter being considered at $O(p_r^4)$, i.e.\ at the level of the fourth power of the eccentricity] are of the form
$A_\textrm{4PN}(u) = \nu( a_5^c + a_5^{\ln} \ln u + \nu a'_5)u^5$,
${\bar D}_\textrm{4PN}(u) =  \nu ( \bar d_4^c + \bar d_4^{\ln} \ln u + \nu \bar d'_4) u^4$, 
and $\hat{Q}_\textrm{4PN}(u,p_r) = \nu q_{4, 3}(\nu)  u^3 p_r^4 + \nu q_{6, 2}(\nu) u^2 p_r^6 + O( p_r^8)$,
with $q_{4, 3}(\nu) = q_{4, 3}  + \nu q'_{4, 3} + \nu^2 q''_{4, 3}$
and $q_{6,2}(\nu) = q_{6,2}  + \nu q'_{6,2} + \nu^2 q''_{6,2}$.
Among the 4PN coefficients, the 1SF contributions are the ones at order $O(\nu)$,
i.e.\ the unprimed ones in our notation, namely
\begin{subequations}
\begin{align}
a_5^c &= \frac{2275\pi^2}{512} - \frac{4237}{60} + \frac{128}{5}\gE + \frac{256}{5}\ln2,
\\[1ex]
a_5^{\ln} &=  \frac{64}{5},
\\[1ex]
 \bar d_4^c  &=  -\frac{533}{45} - \frac{23761\pi^2}{1536} + \frac{1184}{15}\gE
\nonumber\\[1ex]&\qquad
- \frac{6496}{15}\ln2 + \frac{2916}{5}\ln3 ,
\\[1ex]
 \bar d_4^{\ln} &= \frac{592}{15},
\\[1ex]
 q_{4, 3} &=   -\frac{5308}{15} + \frac{496256}{45}\ln2 - \frac{33048}{5}\ln3,
\\[1ex]
 q_{6,2} &= -\frac{827}{3} - \frac{2358912}{25}\ln2
\nonumber\\[1ex]&\qquad
+ \frac{1399437}{50}\ln3
+ \frac{390625}{18}\ln5.
\end{align}
\end{subequations}
The numerical values of the nonlogarithmic coefficients are
\begin{subequations}
\begin{align}
a_5^c &=
23.50338\, 92426\, 03436\, 23875\, 76146\cdots,
\\[1ex]
\bar d_4^c &=
221.57199\, 11921\, 48164\, 02323\, 37716\cdots,
\\[1ex]
q_{4, 3} &=
28.71104\, 42849\, 55949\, 75749\, 69412\cdots,
\\[1ex]
q_{6,2} &=
-2.78300\, 76369\, 52232\, 48902\, 84545\cdots.
\end{align}
\end{subequations}

All these 1SF-order coefficients have been independently checked, either analytically, or  numerically, 
by various SF computations (which used the first law of binary dynamics as a bridge). 
We have already mentioned above the theoretical consensus on the 4PN logarithmic contributions
$a_5^{\ln}$ and $ \bar d_4^{\ln}$. [Note the absence of any logarithmic contribution to  $Q_\textrm{4PN}(u,p_r) $.]
The  nonlogarithmic contribution $a_5^c$ to the main EOB radial potential $A(u)$ is actually not a deep check of 
the ADM results \eqref{SDJS} and \eqref{SDJS'} because  Ref.\ \cite{Damour:2014jta} used the analytical SF result
of Bini and Damour \cite{Bini:2013zaa} to calibrate their single IR ambiguity constant $C$. Let us, however, note two things.
First, all the transcendental contributions to $a_5^c$ have been reproduced by the local ADM computation 
\cite{Jaranowski:2013lca,Jaranowski:2015lha}, the needed calibration of $C$ using only a rational shift.
Second, the analytical determination of $a_5^c$ in  Ref.\ \cite{Bini:2013zaa}
(which was preceded by accurate numerical estimations \cite{Blanchet:2010zd,Barausse:2011dq})
was later analytically confirmed in Refs.\ \cite{Kavanagh:2015lva} and \cite{Johnson-McDaniel:2015vva},
as well as numerically confirmed by extremely high-accuracy self-force results \cite{Shah:2013uya}. 

The nonlogarithmic 1SF 4PN contribution $ \bar d_4^c $ to the second EOB potential $\bar D(u)$ has been fully confirmed
(again using the first law of mechanics) by recent SF works, both analytically and numerically.
As already mentioned, direct analytical checks have been obtained in
Refs.\ \cite{Bini:2015bfb} and \cite{Hopper:2015icj}.
In addition, Ref.\ \cite{Bini:2015bfb} has pointed out that the recent numerical SF results  
of Ref.\ \cite{vandeMeent:2015lxa} provide a numerical check on the value  $ \bar d_4^c $ at the accuracy level $\pm \, 0.05$.
(The latter error level takes into account the fact that the $a_5$ terms have been fully confirmed.)

The  1SF 4PN contributions to $Q(r,p_r)$ have also been recently confirmed, both analytically \cite{Hopper:2015icj} 
and numerically, with an uncertainty $\delta q_{4,3} =\pm 4$ \cite{vandeMeent:2015lxa} (see also the recent numerical determination of the coefficient $q_4(u)$ of $\nu p_r^4$
in the EOB $Q$ potential \cite{Akcay:2015pjz}).

Let us emphasize that the terms proportional to higher powers of $\nu$ in Eq.\ \eqref{nuexpandedH} are much less sensitive than the
terms of order $\nu$ to subtle regularization ambiguities.
Actually, as shown in Refs.\ \cite{Jaranowski:2012eb,Jaranowski:2013lca}
the regularization subtleties are in direct correspondence with the power of $\nu$.
The terms of order $\nu^3$ and $\nu^4$
are not ambiguous at all (and have been independently derived, in the effective field theory approach,
for the relation between energy and orbital frequency for circular orbits, in Ref.\ \cite{Foffa:2012rn}),
while the terms of order $\nu^2$ are delicate, but can be unambiguously derived
when using dimensional regularization for treating the UV divergences \cite{Jaranowski:2013lca}. 
Actually, as explicitly shown in Ref.\ \cite{Bernard:2015njp} (see also the next section),
the recent harmonic-coordinates Fokker-action computation of Ref.\ \cite{Bernard:2015njp} agrees
(modulo some contact transformation) with the action \eqref{SDJS} for all powers of $\nu$,
{\it except} for the first power.
We think that this is related to the fact that the most delicate IR effects are linked with the nonlocal 
contribution \eqref{Snonloc}, which is easily seen to be  purely of order $\nu^1$ [in $S/(M\mu)$]. 
[This is also explicitly displayed in Eqs.\ (7.5)--(7.7) of Ref.\ \cite{Damour:2015isa}.]
As we shall further discuss below, physical effects mixing local and nonlocal effects,
and thereby being sensitive to IR divergences, are very delicate to determine unambiguously.
We can, however, conclude that the non-IR-sensitive part of the results of Ref.\ \cite{Bernard:2015njp}
provide an independent confirmation of all the terms in the action \eqref{SDJS} 
which are of order $\nu^n$ with $n\geq2$.

To summarize this section, many independent SF results have confirmed
all the (IR-sensitive) terms {\it linear in $\nu$},
while the other PN calculations at the 4PN level
(Ref.\ \cite{Foffa:2012rn} and, especially,  Ref.\ \cite{Bernard:2015njp}),
have confirmed all the terms {\it nonlinear in $\nu$}  (i.e.\ $\propto\nu^2$, $\nu^3$, and $\nu^4$).  
We conclude that all the results of Refs.\ \cite{Damour:2014jta,Damour:2015isa} have been (piecewise) confirmed.

\section{Incompatibilities between the 4PN harmonic Fokker action results of Ref.\ \cite{Bernard:2015njp} and self-force results}

The simplest way to compare the results of the recent harmonic Fokker action computation \cite{Bernard:2015njp} to self-force data
is to compute the SF effects induced by the {\it difference} between the dynamics of Ref.\ \cite{Bernard:2015njp}, and that of 
Refs.\ \cite{Damour:2014jta,Damour:2015isa}. This difference has been worked out in Ref.\ \cite{Bernard:2015njp}.

Reference \cite{Bernard:2015njp} obtained, after applying a suitable contact transformation
(alluded to, though not explicitly presented, at the beginning of their Sec.\@~V~B)
to their original harmonic-coordinates result an action of the same form as the ADM results \eqref{SDJS}, namely
\begin{align}
S_\textrm{B$^3$FM} &= \int \Big[\sum_a p_{ai}\md x^i_a
- H^\textrm{loc}_\textrm{B$^3$FM}(\vecxa,\vecpa;s)\md t\Big]
\nonumber\\&\quad
+ S^\textrm{nonloc ($s$)},
\end{align}
with the same nonlocal (or ``tail'') action\footnote{We use here the fact that,
at the 4PN level, $M_{\rm B^3FM} = M_{\rm DJS} + O(1/c^2)$
with $M_{\rm DJS} \equiv M^{\rm here} = m_1 + m_2$.},
but with a different local Hamiltonian,
\begin{align}
\label{eq5}
H^\textrm{loc}_\textrm{B$^3$FM}(\vecxa,\vecpa;s)
&= H^\textrm{loc}_\textrm{B$^3$FM}(\vecxa,\vecpa;r_{12})
\nonumber\\[1ex]&\quad
+ F[\vecxa,\vecpa]\ln\frac{r_{12}}{s}.
\end{align}
(Reference \cite{Bernard:2015njp} does not explicitly display the $\ln s$ dependence of the local Hamiltonian
but agrees with Ref.\ \cite{Damour:2014jta} on the cancellation of that dependence.)
For simplicity, we do not discuss here the issue of the order reduction of the derivatives of $\vecx_a$ and $\vecp_a$ entering (via $I_{ij}^{(3)}$) both the nonlocal action \eqref{Snonloc} and the coefficient $F$ of the logarithm, Eq.\ \eqref{eq6}.
Indeed, there is complete agreement, at the level of the action, between Refs.\ \cite{Damour:2014jta} and \cite{Bernard:2015njp} for what concerns the nonlocal piece of the action.
[As said in Ref.\ \cite{Damour:2014jta}, the order reduction of $I_{ij}^{(3)}$ (i.e.\ its on-shell replacement by a local function of $\vecx_a (t)$ and $\vecp_a (t)$) entails a suitable nonlocal shift of the dynamical variables (which was explicated in Eqs.\ (5.14), (5.15) of Ref.\ \cite{Bernard:2015njp}).]
The important issue is the difference between the two local Hamiltonians which, according to Eq.\ (5.19) of Ref.\ \cite{Bernard:2015njp}, is
\begin{align}
\label{eq7}
H^\textrm{loc}_\textrm{B$^3$FM} - H^\textrm{loc}_\textrm{DJS}
&= \frac{G^4 M m_1^2 m_2^2}{c^8 r_{12}^4}
\nonumber\\[1ex]&\quad
\times \Bigg[a_\textrm{B$^3$FM}\Big(\frac{\npi}{m_1}-\frac{\npii}{m_2}\Big)^2
\nonumber\\[1ex]&\quad
+ b_\textrm{B$^3$FM}\Big(\frac{\vecp_1}{m_1}-\frac{\vecp_2}{m_2}\Big)^2
+ c_\textrm{B$^3$FM} \frac{GM}{r_{12}}\Bigg],
\end{align}
with
\be
\label{eq8}
(a,b,c)_\textrm{B$^3$FM}
= \left(\frac{1429}{315},\frac{826}{315},\frac{902}{315}\right).
\ee

Before discussing further the origin of the discrepancy
[i.e.\ the fact that $(a,b,c)_{\rm B^3FM}\ne(0,0,0)$],
we wish to emphasize two things: (i)  the discrepant terms in Eq.\ \eqref{eq7}
represent only three (Galileo-invariant) terms among the hundreds (exactly 219)
of contributions to the (non-center-of-mass) two-body Hamiltonian,
as displayed in the Appendix of Ref.\ \cite{Damour:2014jta},
or in Sec.\ VIII E of Ref.\ \cite{Jaranowski:2015lha};
(ii) the three discrepant terms \eqref{eq7} 
[when expressed in the center of mass, and in reduced variables, i.e.\ in the sense of Eq.\ \eqref{nuexpandedH}]
are {\it linear} in the symmetric mass ratio  $\nu$, i.e.\ they are of 1SF order.

The discrepancy  Eq.\ \eqref{eq7} therefore implies 1SF-detectable effects away from all the 1SF checks of the ADM
dynamics \eqref{SDJS} reviewed in the previous section. Let us now quantify the corresponding 1SF differences.
To do that it is convenient to transcribe the 4PN-level Hamiltonian difference  \eqref{eq7}  in terms of corresponding 
differences in the three EOB potentials $A$, $\bar D$, $Q$.
(We recall in passing that the EOB parametrization of the dynamics in terms of $A$, $\bar D$, $Q$ is completely gauge fixed,
and therefore directly linked to gauge-invariant quantities.)

Considering, for more generality, a 4PN-level Hamiltonian difference with general coefficients $(a,b,c)$ in Eq.\ \eqref{eq7}, the
corresponding additional contributions to $A$, $\bar D$, $Q$ are easily found to be 
\begin{subequations}
\label{eq9+10}
\begin{align}
\label{eq9}
\delta^{a,b,c} A &=( 2 b + 2 c) \, \nu \, u^5 ,
\\[1ex]
\label{eq10}
\delta^{a,b,c}\bar{D} &=( 2 a + 8 b)  \, \nu \, u^4 ,
\\[1ex]
\delta^{a,b,c} Q &= 0 .
\end{align}
\end{subequations}
Note in passing that those contributions are invariant under the ``gauge transformation''
\be
\label{eq11}
\delta^g (a,b,c) = g(4,-1,1),
\ee
which corresponds to the most general canonical transformation respecting the structure \eqref{eq7} (with generating function $\propto g \, \nu \, p_r/r^3$).
Inserting the values \eqref{eq8} in Eqs.\ \eqref{eq9}--\eqref{eq10} leads to changes in $A$ and $\bar{D}$ (with respect to the values computed in Ref.\ \cite{Damour:2015isa})
\begin{align}
\label{deltaA}
\delta^\textrm{B$^3$FM}_\textrm{4PN} A &= \frac{384}{35}\,\nu\,u^5 ,
\\[1ex]
\label{deltaD}
\delta^\textrm{B$^3$FM}_\textrm{4PN} \bar{D} &= \frac{9466}{315}\,\nu\,u^4 .
\end{align}

In other words, the only 1SF 4PN-level coefficients that are different are $a_5^c$ and $\bar d_4^c$ with
\begin{align}
\label{deltaa5}
\delta^\textrm{B$^3$FM} a_5^c &= \frac{384}{35} \approx 10.97143 ,
\\[1ex]
\label{deltad4}
\delta^\textrm{B$^3$FM} \bar{d}_4^c &= \frac{9466}{315} \approx 30.05079 .
\end{align}

Such large, 4PN-level deviations away from the results of Ref.\ \cite{Damour:2015isa}
are in violent contradiction with the many SF confirmations of the ADM 4PN dynamics reviewed in the previous section.
They have been obtained here by using the Delaunay-like reduction used in Ref.\ \cite{Damour:2015isa} to convert the nonlocal dynamics
\eqref{SDJS} into the (formally) local one  \eqref{SDJS'}.  The authors of Ref.\ \cite{Bernard:2015njp} express doubts about some aspects
of the results of \cite{Damour:2015isa}.
We shall address their concerns in the following sections and conclude that their concerns are unsubstantiated.
Therefore we conclude that taking the results of Ref.\ \cite{Bernard:2015njp} at face value
does lead to the large changes \eqref{deltaa5} and \eqref{deltad4} above, and are therefore strictly incompatible with extant SF knowledge.

We wish to go further and point out an even more blatant contradiction with SF tests of periapsis precession.
First, we note (using, e.g.\ Eq.\ (8.3) in \cite{Damour:2015isa}) that the changes above entail a corresponding change
in the 4PN-level precession coefficient
$\rho_4^c$ given by $\delta\rho_4^c= 10 \delta a_5^c + \delta \bar d_4^c$, i.e.
\be
\delta^\textrm{B$^3$FM} \rho_4^c = (2 a + 28 b + 20 c )_\textrm{B$^3$FM},
\ee
i.e.
\be
\label{deltarho4}
\delta^\textrm{B$^3$FM} \rho_4^c = \frac{44026}{315} \approx 139.7650794.
\ee
As we recalled above, SF computations \cite{Barack:2010ny,vandeMeent:2016} of the precession function $\rho(u)$ \cite{Damour:2009sm},
and of its PN coefficients, do not rely on the first law of binary mechanics, but involve direct computations of the additional precession induced by the nonlocal self-force.
The difference \eqref{deltarho4} is therefore (assuming the correctness of the Delaunay reduction of Ref.\ \cite{Damour:2015isa})
a direct dynamical consequence of the Fokker-action result of Ref.\ \cite{Bernard:2015njp}.
It would be interesting to confirm this result by a purely dynamical computation of the (nonlocal) 4PN precession.
We note here that the difference is already excluded by the ``old" result \eqref{drho4barack},
and even more so (by $10^8$ standard deviations!) by the recent one \eqref{drho4meent}.

\section{Logical basis of the action-angle Delaunay-like method of Ref.\ \cite{Damour:2015isa}}

Independently of the above SF confirmations of the 4PN results of Refs.\ \cite{Damour:2014jta,Damour:2015isa},
let us reassess the logical basis and the consistency of the methodology used in our Delaunay-EOB derivation,
and indicate, by contrast, where, in our opinion, lie the flaws of Ref.\ \cite{Bernard:2015njp} that have led to the nonzero values \eqref{eq8},
which are incompatible with many SF results. 
Our Ref.\ framework is the action-angle formulation of (planar) Hamiltonian dynamics.
In \cite{Damour:2015isa} we used the standard Delaunay notation (modulo the use of calligraphic letters $L \to {\mathcal L}$, $G \to {\mathcal G}$). Namely, $({\mathcal L} , \ell ; {\mathcal G} , g)$ are the two planar action-angle canonical pairs: ${\mathcal L} = \sqrt a$ is conjugate to the mean anomaly angle $\ell$, while ${\mathcal G} = \sqrt{a(1-e^2)}$ is conjugate to the argument of the periastron $g = \omega$. In order to better exhibit the meaning (and consistency) of our approach in the circular limit,
we shall use here the combination of Delaunay variables introduced by Poincar\'e, namely, the two action-angle pairs $(\Lambda,\lambda;I_r,\varpi)$
where (using suitably scaled variables as in Ref.\ \cite{Damour:2015isa})
\begin{subequations}
\begin{align}
\Lambda &= {\mathcal L} = I_r + I_{\varphi},
\\[1ex]
\lambda &= \ell + g,
\\[1ex]
I_r &= {\mathcal L} - {\mathcal G} = {\mathcal L} - I_{\varphi},
\\[1ex]
\varpi &= -g = -\omega.
\end{align}
\end{subequations}
Here, $I_{\varphi} = {\mathcal G}$ is the angular momentum
($I_{\varphi} = \frac1{2\pi} \oint p_{\varphi} \, \md\varphi = p_{\varphi}$)
and $I_r$ is the radial action ($I_r = \frac1{2\pi} \oint p_r \, \md r$).
The sum $\Lambda = I_r + I_{\varphi}$ is conjugate to the mean longitude $\lambda = \ell + g$, while $I_r$ is conjugate to (minus) the argument of the periastron $\varpi = -\omega$. The latter (surprising) minus sign is necessary to have
$\md L \wedge \md\ell + \md G \wedge \md g = \md \Lambda \wedge \md\lambda + \md I_r \wedge \md\varpi$
with $I_r = {\mathcal L} - {\mathcal G} = \sqrt a - \sqrt{a(1-e^2)} > 0$. In the circular limit, $\lambda$ becomes the usual polar angle $\varphi$ in the orbital plane and $I_r \to 0$, so that the action variable $\Lambda$ becomes equal to the angular momentum $I_{\varphi} = p_{\varphi}$.  In this limit, the general Hamilton equation $\dot\lambda = \partial H / \partial\Lambda$ gives back the usual circular link between the orbital frequency $\Omega = \dot\varphi$ and the derivative of the energy with respect to the angular momentum $p_{\varphi}$. However, when dealing with noncircular orbits, and tackling the nonlocal action \eqref{Snonloc}, it is important to work with the clearly defined canonical action-angle pairs $(\Lambda , \lambda ; I_r , \varpi)$.
In principle, it would be better, when discussing the circular limit,
to replace the second pair $(I_r , \varpi)$ by the associated Poincar\'e variables
$\xi = \sqrt{2I_r} \cos \varpi$, $\eta = \sqrt{2I_r} \sin \varpi$,
because they are canonical ($\md\xi \wedge \md\eta = \md I_r \wedge \md\varpi$)
and regular in the circular limit (while $\varpi$ becomes ill defined as $I_r \to 0$).
Keeping in mind such an additional change of variables,
we shall, however, find simpler to express our methodology in terms of $I_r$ and $\varpi$.

The basic methodology we used in Ref.\ \cite{Damour:2015isa} for dealing with the nonlocal 4PN dynamics, Eqs.\ \eqref{SDJS} and \eqref{Snonloc}, consists of four steps:
(a) we reexpress the action \eqref{SDJS} as a nonlocal action in the action-angle variables $(\Lambda , \lambda ; I_r , \varpi)$;
(b) we expand it (formally to infinite order) in powers of the eccentricity, i.e.\ in powers of $\sqrt{I_r}$;
(c) we ``order reduce'' the nonlocal dependence on the action-angle variables by using the on-shell equations of motion\footnote{As discussed in detail below, this is equivalent to applying suitable nonlocal shifts of the phase-space variables.};
and (d) we eliminate, \textit{\`a la} Delaunay, the periodic terms in the order-reduced Hamiltonian by a canonical transformation of the action-angle variables. After these four steps, we end up with an Hamiltonian which is an ordinary (local) function of the (transformed) action variables alone.

In order not to get distracted by irrelevant technicalities, let us illustrate this methodology
[which was applied in Ref.\ \cite{Damour:2015isa} to the full 4PN action \eqref{SDJS}--\eqref{Snonloc}]
on a simpler toy example which contains some of the key ingredients of the action \eqref{SDJS}--\eqref{Snonloc}, namely the action
\begin{align}
S_\textrm{toy} &= \int[\vecp(t)\cdot\md\vecr(t) - H_\textrm{toy}(t)\md t],
\\[1ex]
H_\textrm{toy}(t) &= \frac{1}{2}\vecp(t)^2 - \frac{1}{r(t)}
\nonumber\\[1ex]&\quad
+ \varepsilon \int_{-\infty}^{+\infty}\md t' \mu(t' - t)
\frac{\vecr(t')\cdot\vecr(t)}{|\vecr(t')|^3|\vecr(t)|^3},
\end{align}
where $\mu (\tau)$ is an even function of the time difference $\tau \equiv t'-t$.
In the real case \eqref{SDJS}--\eqref{Snonloc}, $\mu (\tau) = 1/\vert \tau \vert$
with an additional partie finie ($\Pf$) prescription.
The subtleties linked to the $\Pf$ prescription and to the slow decay of $1/\vert\tau\vert$
for $\vert\tau\vert\to+\infty$ are only of a technical nature.
For instance, the oscillatory nature of the integrand $I'^{(3)}_{ij} I_{ij}^{(3)}$ [mimicked as $x'_i \, x_i/(r'^3 r^3)$ in our toy example]
ensures the large $\vert \tau \vert$ convergence\footnote{We have {\it convergence} both at $\tau \to +\infty$ and $\tau \to -\infty$, without having absolute convergence.
See Appendix A for more discussion of convergence issues in reduced actions.}
of the nonlocal integral \eqref{Snonloc} without having to assume that $I_{ij}^{(3)}$ tends to zero when $\vert \tau \vert \to + \infty$.
[Indeed, the fact that $- x_i/r^3$ is the on-shell value of the second time derivative of $ x_i$,
and $I_{ij}^{(3)}$ is the third time derivative of $I_{ij}$, ensures that their large-time averages both vanish for conservative bound motions.
Note that our toy model is the ``electromagnetic" (dipolar) analog of the gravitational (quadrupolar) tail action.]
For simplicity, as we wish here to emphasize issues of principle without getting bogged down by secondary technical issues,
we will not specify the weight $\mu(\tau)$ used in our toy model, but proceed as if it were a smooth, integrable {\it even} function of $\tau$.

The first two steps, (a) and (b), of our procedure yield
(when considered, for illustration, at linear order in the eccentricity $e$)
\begin{align}
H_\textrm{toy}(t) &= -\frac{1}{2\Lambda^2} + \varepsilon\int_{-\infty}^{+\infty}\md\tau\,\mu(\tau)
\frac{1}{\Lambda^4\Lambda'^4}\Big[\cos(\lambda'-\lambda)
\nonumber\\[1ex]&\quad
+ 2 e \cos(2\lambda-\lambda'+\varpi) + 2 e' \cos(2\lambda'-\lambda+\varpi')
\nonumber\\[1ex]&\quad
+ O(e^2)\Big].
\end{align}
Here and below $f'$ denotes the variable $f$ taken at time $t' = t+\tau$,
while $f$ denotes its value at time $t$.

Step (c) consists in writing, for any variable $f$,
its value $f'$ at the shifted time $t'$ in terms of canonical variables at time $t$
and of an integral over intermediate times involving the rhs of the (Hamiltonian) equations of motion.
For a nonlocal action, the latter read ${S}_{\lambda}(t) = 0$,
${S}_{\Lambda}(t) = 0$, etc., with
\begin{subequations}
\begin{align}
\label{eoma}
{S}_{\lambda}(t) &\equiv \frac{\delta S}{\delta \lambda(t)} \equiv - \dot{\Lambda}(t) - \frac{\delta H}{\delta\lambda(t)},
\\[1ex]
\label{eomb}
{S}_{\Lambda}(t) &\equiv \frac{\delta S}{\delta \Lambda(t)} \equiv \dot{\lambda}(t) - \frac{\delta H}{\delta\Lambda(t)},
\quad \text{etc.},
\end{align}
\end{subequations}
where $\delta/\delta f(t)$ denotes a {\it functional} derivative
[acting on the action $S$ or on $\int\md t\,H^{\rm nonlocal}(t)$].
In the normal case of local actions \cite{Schafer:1984mr,Damour:1985mt},
the use of the identities \eqref{eoma} and \eqref{eomb} is the main tool allowing
one to show how the replacement of the equations of motion within an action is equivalent to a suitable shift of the dynamical variables (a ``field redefinition").
In the case of nonlocal actions, one must {\it integrate} the identities \eqref{eoma} and \eqref{eomb}
before replacing  $\lambda'\equiv\lambda(t+\tau),\Lambda'\equiv\Lambda(t+\tau),\cdots$ in the nonlocal piece of the action. 
In this integration, the terms ${S}_{\lambda}(t), {S}_{\Lambda}(t), \cdots$
are treated as additional terms, which would be zero on shell, but which are now considered as ``source terms" on the rhs of the
usual Hamilton equations of motion, $\dot{\lambda}(t) - {\delta H}/{\delta\Lambda(t)}=S_{\Lambda}, \cdots$.
In view of the well-known possibility of neglecting ``double-zero terms,'' it is enough to work {\it linearly} in these (simple-zero) source terms.
Moreover, as we are working within a PN-expanded scheme (and as the nonlocality enters only at order $\varepsilon = 1/c^8$),
it is actually enough [modulo terms of order $O(c^{-2} \varepsilon)$], for the purpose of 
replacing $\lambda', \Lambda', \cdots$ in the nonlocal piece of the action, to compute them as solutions 
of  the (sourced) {\it Newtonian-level} equations of motion, i.e.\ solutions of  the differential equations \eqref{eoma} and \eqref{eomb}
with $\delta H / \delta\Lambda, \cdots \to \delta H_0 / \delta\Lambda, \cdots$, where $H_0=-1/(2\Lambda^2)$.
The explicit form of these sourced equations of motion are
\begin{subequations}
\begin{align}
\label{eoma'}
\dot{\Lambda}(t) &= -{S}_{\lambda}(t) ,
\\[2ex]
\label{eomb'}
\dot{\lambda}(t) &= \frac1{\Lambda^3(t)}
+ {S}_{\Lambda}(t), \quad \text{etc.}
\end{align}
\end{subequations}
Viewing these equations as differential equations with respect to
$t'=t+\tau$ (or $\tau$) for the unknowns $\Lambda' \equiv \Lambda(t')$
and $\lambda' \equiv \lambda(t')$, and imposing the initial conditions
at $t'=t$ (i.e.\ $\tau=0$), $\Lambda'(t'=t)=\Lambda$, $\lambda'(t'=t)=\lambda$,
yields as a unique solution (to linear order in the source terms)
\begin{subequations}
\label{eqs23-24}
\begin{align}
\Lambda' & = \Lambda - \int_t^{t'} \md t_1 S_{\lambda}(t_1), 
\\[1ex]
\lambda' & = \lambda + \frac{t'-t}{\Lambda^3} + \frac{3}{\Lambda^4} \int_t^{t'} \md t_1 (t'-t_1) S_{\lambda} (t_1)
\nonumber\\[1ex]&\quad
+ \int_t^{t'} \md t_1 S_{\Lambda} (t_1).
\end{align}
\end{subequations}
(We do not assume here $t<t'$ but use the convention that $\int_t^{t'} = - \int_{t'}^{t}$.)
When replacing the identities \eqref{eqs23-24} in a nonlocal action
$S=\int\md t\,\md t'\mu(t'-t)\,{\cal S}(\Lambda,\Lambda',\lambda'-\lambda,\cdots)$
(keeping only the terms linear in the source terms, which is allowed modulo ``double-zero" terms),
the extra terms involving the source contributions will have the form 
$\int\md t_1 \, \left[ \xi_{\Lambda}( t_1) \, S_{\Lambda} (t_1)+ \xi_{\lambda}( t_1) \, S_{\lambda} (t_1) + \cdots \right] $, 
where the quantities $ \xi_{\Lambda}( t_1) , \xi_{\lambda}( t_1), \cdots$ are given by  integrals over $t$ and $t'$ 
of a function of dynamical variables: 
$\xi_{\Lambda}( t_1) = \iint \md t\,\md t' \mu(t'-t) \mathcal{A}(t'-t, t_1 -t,\Lambda(t),\lambda(t))$, $\cdots$.
(Note that, because of the inequalities  $t < t_1 < t'$ or  $t' < t_1 < t$,
for given values of $t_1$ and $\tau=t'-t$, $t$ and $t'$ both range over a bounded interval.)
The  extra terms involving the source contributions can then be 
``field-redefined away'' by  corresponding  shifts of the dynamical variables:
$\delta \Lambda(t_1) = \xi_{\Lambda}(t_1), \delta \lambda(t_1) = \xi_{\lambda}(t_1), \cdots$.
[These shifts have, in general, nonlocal structures of the type
$\delta\Lambda(t_1) = \int\md t\, K_{\Lambda}(t-t_1,\Lambda(t),\lambda(t))$,
$\delta\lambda(t_1) = \int\md t\, K_{\lambda}(t-t_1,\Lambda(t),\lambda(t))$, etc.]

After performing these (nonlocal) shifts, we are left with an action
obtained by replacing $\lambda', \Lambda', \cdots$ by the rhs of Eq.\ \eqref{eqs23-24}
in which one has set to zero the terms ${S}_{\lambda}(t_1), {S}_{\Lambda}(t_1),\cdots$.
In other words, the use of these shifts justifies the naive replacement of 
$\lambda', \Lambda', \cdots$ by the solution of the (Newtonian) equations of motion, namely
 \begin{subequations}
\begin{align}
\label{repla}
\lambda' &= \lambda + \frac\tau{\Lambda^3},
\\[1ex]
\label{replb}
\Lambda' &= \Lambda.
\end{align}
\end{subequations}
Note that the very simple time structure of the unperturbed solution in action-angle variables
plays a very useful role in our Delaunay-based reduction procedure.

With the replacements \eqref{repla} and \eqref{replb} [i.e.\ after steps (a), (b), and (c)]
we have a Hamiltonian of the form
\begin{align}
\label{Htoy1}
H'_\textrm{toy}(t) &= -\frac{1}{2\Lambda^2} + \varepsilon \int_{-\infty}^{+\infty}\md\tau\,\mu(\tau)
\frac{1}{\Lambda^8}\bigg\{\cos\left(\frac{\tau}{\Lambda^3}\right)
\nonumber\\[1ex]&\quad
+ 2 e \bigg[ \cos\left(\frac{\tau}{\Lambda^3}-\lambda-\varpi\right)
\nonumber\\[1ex]&\quad
+ \cos\left(\frac{2\tau}{\Lambda^3}+\lambda+\varpi\right) \bigg]
+ O(e^2)\bigg\},
\end{align}
where we recall that $e$ is the function of $I_r / \Lambda$
defined so that $1 - \sqrt{1-e^2} = I_r / \Lambda$, i.e.
\be
e = \sqrt{1 - \left( 1- \frac{I_r}\Lambda \right)^2}.
\ee

At this stage, Eq.\ \eqref{Htoy1} defines an ordinary Hamiltonian,
expressed in terms of action-angle variables $(\Lambda,\lambda;I_r,\varpi)$.
[Strictly speaking, the variables $\Lambda$, $\lambda$, etc., at the stage of Eq.\ \eqref{Htoy1}
differ from the ones entering the original action
by the shifts $\xi_{\Lambda}$, $\xi_{\lambda}$, etc., mentioned above.
They should be denoted as $\Lambda^{\rm shifted}$, $\lambda^{\rm shifted}$, etc.,
but, for simplicity, we do not indicate the shifted nature of $\Lambda$, $\lambda$, $\ldots$\ .]

As we are now dealing with an ordinary, local Hamiltonian, our fourth step (d),
i.e.\ the Delaunay elimination of periodic terms by suitable canonical transformations is standard.
For instance, a generating function proportional to
\be
\varepsilon \int_{-\infty}^{+\infty}\md\tau\,\mu(\tau) \frac{1}{\Lambda^5}
\sin\left(\frac{\tau}{\Lambda^3}-\lambda-\varpi\right)
\ee
will eliminate the contribution $+2e\cos(\tau/\Lambda^3-\lambda-\varpi)$
on the rhs of Eq.\ \eqref{Htoy1}.

After completing step (d) (formally to all orders in the eccentricity)
we end up with an ordinary (local) Hamiltonian that depends {\it only on (shifted) action variables}.
E.g., for our toy model
\begin{align}
\label{Htoy2}
H''_\textrm{toy}(\Lambda,I_r) &= -\frac{1}{2\Lambda^2} + \varepsilon \int_{-\infty}^{+\infty}\md\tau\,\mu(\tau)
\frac{1}{\Lambda^8}
\nonumber\\[1ex]&\qquad
\times\bigg[\cos\left(\frac{\tau}{\Lambda^3}\right)
+ O(e^2)\bigg].
\end{align}

Let us emphasize that our procedure allows one, in particular,
to construct (formally to {\it all} orders in eccentricity)
a {\it strictly conserved energy} that consistently includes the nonlocal tails at the 4PN level.
[The nonlocal effects are recognizable through their integral nature, e.g. the
integral $ \varepsilon \int_{-\infty}^{+\infty}\md\tau\,\mu(\tau) (\cdots)$ in Eq.\ \eqref{Htoy2}.]
Namely, our final, Delaunay Hamiltonian is strictly conserved.
This shows that the contrary statement below Eq.\ (5.17) in Ref.\ \cite{Bernard:2015njp}
is incorrect (at least when allowing for an expansion in powers of $I_r/\Lambda \sim e^2$).

In addition, our construction also gives two conserved action variables\footnote{
It would be interesting to study whether our (4PN-order) construction of conserved action variables, and conserved energy,
is related to the (1SF-order) work of Ref.\ \cite{Vines:2015efa}.}: $\Lambda$ and $I_r$.
These conserved quantities are such that the difference,
$\Lambda - I_r$ is conserved and reduces to the usual orbital angular momentum $I_{\varphi} = \frac1{2\pi} \oint p_{\varphi} \,\md\varphi = p_{\varphi}$ in the local case.
We can then consider that $I_{\varphi} := \Lambda - I_r$ defines a {\it strictly conserved angular momentum} that consistently includes the 4PN nonlocal dynamical effects.
(Indeed, one easily checks that the rotational symmetry $\varphi' = \varphi + {\rm const}$ remains a symmetry at all stages of our construction
so that $I_{\varphi}$ is indeed the Noether conserved quantity associated with this symmetry.)

Let us also emphasize another consistency feature of our final (reduced, local, Delaunay) Hamiltonian. As the above detailed derivation clearly shows, our final Hamiltonian action
(in terms of $\Lambda^{\rm shifted}$, $\lambda^{\rm shifted}$, \ldots
but without explicitly indicating the shifts) is simply
\be
S'' = \int\big[\Lambda\md\lambda + I_r\md\varpi - H''(\Lambda,I_r)\md t\big],
\ee
and the corresponding equations of motion are the usual ones:
\be
\dot{\lambda} = \frac{\partial H''(\Lambda,I_r)}{\partial\Lambda},
\quad
\dot{\varpi} = \frac{\partial H''(\Lambda,I_r)}{\partial I_r},
\quad \dot{\Lambda} = 0,
\quad \dot{I}_r = 0.
\ee
We said above that, in the circular limit, the mean anomaly
$\lambda = \ell + g = \ell + \omega$ reduces to a usual polar angle $\varphi$.
(Indeed, this is true in the Newtonian case, and,
as all the Delaunay shifts are periodic in the unique {\it fast} angular variable $\lambda$,
we have that $\varphi = \lambda \, +$ terms periodic in $\lambda$ that vanish with $e$.)
Therefore our derivation shows, among other things,
that the orbital frequency $\Omega$ $(=\dot\lambda)$ of a circular $(I_r=0)$ binary
(with conserved angular momentum $\Lambda = I_{\varphi}$) is given by the usual formula
\be
\Omega = \dot{\lambda} = \left.\frac{\partial H''(\Lambda,I_r)}{\partial\Lambda}\right|_{I_r=0}
= \left.\frac{\partial H''(I_r+I_\varphi,I_r)}{\partial I_\varphi}\right|_{I_r=0}.
\ee
In particular, this means, in our toy example, that the circular angular frequency is of the form
\begin{align}
\Omega_\textrm{toy} &= \frac{\partial H''_\textrm{toy}(\Lambda,0)}{\partial\Lambda}
\nonumber\\[1ex]&\quad
= \frac{1}{\Lambda^3} + \varepsilon\,\frac{\partial}{\partial\Lambda}
\left(\int_{-\infty}^{+\infty}\md\tau\,\mu(\tau)
\frac{1}{\Lambda^8}\cos\left(\frac{\tau}{\Lambda^3}\right)\right).
\end{align}
One sees that the computation of the rhs will involve two types of contributions
coming from the tail term: a ``normal'' contribution where the $\Lambda$ differentiation
of the second term acts on the prefactor $1/\Lambda^8$ of the tail integral,
and a ``new'' contribution where $\partial/\partial\Lambda$ acts
on the argument of $\cos (\tau / \Lambda^3)$.

We recall that, in the full 4PN case, the analog of our final result \eqref{Htoy2}
[say, for simplicity, at the $O(e^0)$ level]
is the one discussed in Sec.\ V of Ref.\ \cite{Damour:2014jta},
which leads to (with $J$ now denoting $I_{\varphi} \equiv \Lambda - I_r$)
\begin{align}
E^\textrm{circ}_\textrm{4PN}(J) &= H^\textrm{loc}_\textrm{DJS}(\vecr,\vecp;s)
+ F(\vecr,\vecp)\ln\frac{r}{s}
\nonumber\\[1ex]&\quad
- \frac{1}{5} \frac{G^2M}{c^8} \Pf_{2s/c}\int_{-\infty}^{+\infty}\frac{\md\tau}{|\tau|}f(\tau),
\end{align}
where $f(\tau):=\big[I_{ij}^{(3)}(t+\tau)\,I_{ij}^{(3)}(t)\big]^{\rm circ}$,
as in Sec.\ V of Ref.\ \cite{Damour:2014jta}.
[In the circular limit $f(\tau)$ does not depend on $t$.]
The explicit value of the last, tail term\footnote{
Here, we find it convenient to work with a tail term defined with some given, constant scale $s$.
Such a scale is not a dynamical variable and is not affected by the $\Lambda$ differentiation above,
i.e.\ the $f$ differentiation in the present circular case.}
is given, in view of Eqs.\ (5.7)--(5.9) in Ref.\ \cite{Damour:2014jta} by
\begin{align}
\label{circHtail1}
[H^\textrm{tail ($s$)}]^\textrm{circ}(J) &= \frac{64}{5} \frac{G^2M}{c^8}
(\mu\Omega^3 r_{12}^2)^2 \ln\left(4\me^\gE\frac{\Omega(J)s}{c} \right)
\nonumber\\[2ex]
&= \frac{64}{5} \mu c^2 \frac{\nu}{j^{10}} \ln\left(\frac{4\me^\gE \hat s}{j^3} \right),
\end{align}
where $j \equiv cJ/(GM\mu)$, and $\hat s \equiv s/m$ with $m \equiv GM/c^2$.

If we alternatively decide to define the tail contribution to the reduced Hamiltonian
by incorporating the term $F\ln(r/s)$ in it, we must simply replace the scale $s$ by $r=r_{12}$,
so that (using the fact that along Newtonian circular orbits $r^{\rm circ} = m j^2$)
\be
\label{circHtail2}
[H^\textrm{tail ($s=r_{12}$)}]^\textrm{circ}(J)
= \frac{64}{5} \mu c^2 \frac{\nu}{j^{10}} \ln\left(\frac{4\me^\gE}{j} \right).
\ee
Evidently, we must keep in mind that the meaning of the two quantities \eqref{circHtail1} and \eqref{circHtail2} is different
and that their $J$ derivative will differ by a term involving the $J$ derivative of $r^{\rm circ}=m \, j^2$.
But the latter difference has a conceptually trivial origin [in view of the {\it local} nature of the $F\ln(r/s)$ term in the Hamiltonian]
and does not interfere with the issue we wish to emphasize here,
namely the existence, in the orbital frequency, $\Omega^{\rm circ}=\md H^{\rm circ}/\md J$,
of a contribution coming from differentiating the $J$ dependence
of the argument $4\me^\gE\,\Omega(J)s/c$ of the logarithm in the first (or second) line of Eq.\ \eqref{circHtail1}.

As clearly pointed out already in Ref.\ \cite{Damour:2014jta},
the computation of the tail contribution to the (circular, Delaunay) function $H(J) = E(J)$
``involves the evaluation of the nonlocal piece $\left[ H_{\rm 4PN}^{{\rm tail}\,(s)}\right]$
along circular motion (without any differentiation)''.
This lack of any differentiation in the evaluation of $\left[ H^{\rm tail} \right]^{\rm circ}(J)$
distinguishes it among the other gauge-invariant functions one can associate
with the sequence of circular orbits,
and rendered very clear\footnote{In view, e.g., of the result of Ref.\ \cite{Damour:1999cr}
about the consistency of the order reduction of higher-order Hamiltonians,
see also Sec.\ III in Ref.\ \cite{Damour:2015isa} and below.}
that this necessarily led to the correct value of the function
$\left[H^{\rm tail}\right]^{\rm circ}(J)$. Our discussion above has confirmed
the consistency of this result and has shown that, indeed,
when computing the orbital frequency it is correct to $J$-differentiate the argument of the tail log.

Let us now derive the tail-log contribution to the functional relation between
energy and orbital frequency, $E(\Omega)$, corresponding to Eq.\ \eqref{circHtail2}.

We start by deriving a useful general result about  the function $E(\Omega)$,
or equivalently the function $E(x)$ [where $x \equiv (GM \Omega / c^3)^{2/3}$].
Let us first consider the case of an ordinary Hamiltonian dynamics of the type (with $J\equiv p_\varphi$)
\be
\label{eq39}
H^\textrm{ord}(r,p_r,J) = H_0(r,p_r,J) + \varepsilon H_1(r,p_r,J).
\ee
When considering the sequence of circular orbits, it is well known\footnote{
The on-shell vanishing of $\partial H^\textrm{ord}/\partial r$
ensures that when replacing the exact solution
$r_\textrm{exact}^\textrm{circ}(J)=r_0^\textrm{circ}(J)+\varepsilon r_1^\textrm{circ}(J)$,
the $\varepsilon r_1^\textrm{circ}(J)$ perturbation will only contribute
at order $O(\varepsilon^2)$.}
that, modulo $O(\varepsilon^2)$, one can replace from the start the radius $r$ on the rhs by its \emph{unperturbed} circular value as a function of $J$,
$r_0^\textrm{circ}(J)$, defined as the solution of $\partial H_0(r,p_r=0,J)/\partial r = 0$ for a given $J$.
(We are assuming here that the $p_r$ dependence of $H^\textrm{ord}$ starts at order $p_r^2$
so that one can consistently set $p_r=0$ before considering the $J$ dependence.)
This yields for the circular-reduced dependence of $H^\textrm{ord}$ as a function of $J$
\be
\label{eq40}
H_\textrm{circ}^\textrm{ord}(J) = H_0^\textrm{circ}(J)
+ \varepsilon H_1^\textrm{circ}(J) + O(\varepsilon^2),
\ee
where $H_0^\textrm{circ}(J)=H_0(r_0^\textrm{circ}(J),p_r=0,J)$
and $H_1^\textrm{circ}(J)=H_1(r_0^\textrm{circ}(J),p_r=0,J)$.
Note that the $O(\varepsilon)$ contribution to $H_\textrm{circ}^\textrm{ord}(J)$
is directly equal to the zeroth-order-circular value of the original
$O(\varepsilon)$ contribution to the Hamiltonian \eqref{eq39}
[without any contribution from $\varepsilon r_1^\textrm{circ}(J)$,
and without any differentiation].
Furthermore, the on-shell vanishing (along exact circular orbits)
of $\partial H^\textrm{ord}/\partial r$ also ensures
that the circular-reduced $J$ dependence of the orbital frequency
$\Omega=\partial H^\textrm{ord}(r,p_r,J)/\partial J$ is simply given
by $J$ differentiation of $H_\textrm{circ}^\textrm{ord}(J)$:
\be
\label{eq41}
\Omega^\textrm{circ}(J) = \frac{\md H_\textrm{circ}^\textrm{ord}(J)}{\md J}
= \Omega_0^\textrm{circ}(J) + \varepsilon \frac{\md H_1^\textrm{circ}(J)}{\md J} + O(\varepsilon^2),
\ee
where $\Omega_0^\textrm{circ}(J)=\md H_0^\textrm{circ}(J)/\md J$ is its unperturbed value.

From the two results \eqref{eq40} and \eqref{eq41}, it is easy to derive the perturbed value of the function
$E^\Omega(\Omega)\equiv H_\textrm{circ}^\textrm{ord}(J^\textrm{circ}(\Omega))$,
where $J^\textrm{circ}(\Omega)$ is the inverse of the function $\Omega^\textrm{circ}(J)$.
One gets
\be
\label{eq42}
E^\Omega(\Omega) = E_0^\Omega(\Omega) + \varepsilon E_1^\Omega(\Omega)
+ O(\varepsilon^2),
\ee
with
\be
\label{eq43}
E_1^\Omega(\Omega) = \left[ H_1^\textrm{circ}(J)
- \frac{\md E_0^\Omega}{\md\Omega} \frac{\md H_1^\textrm{circ}(J)}{\md J}
\right]_{J=J_0^\textrm{circ}(\Omega)},
\ee
where $J_0^\textrm{circ}(\Omega)$ is the inverse of the function $\Omega_0^\textrm{circ}(J)$.

Let us apply the convenient result \eqref{eq43} to the $r_{12}$-scale tail perturbation
written as
\be
\label{eq44}
\varepsilon H_1^\textrm{tail ($r_{12}$)}
= \frac{64}{5} \mu c^2 \frac{\nu}{j^{10}}
\ln\left(\frac{4\me^\gE\Omega\, r_0^\textrm{circ}(J)}{c} \right),
\ee
where we leave open for the moment the functional dependence of $\Omega$
within the tail logarithm.
In our case $E_0^\Omega=-\frac{1}{2} \mu c^2 \widehat{\Omega}^{2/3}$
and $j_0^\textrm{circ}(\Omega)=\widehat{\Omega}^{-1/3}$ (where $\widehat{\Omega} \equiv m \, \Omega$),
so that the general result \eqref{eq43} yields the simple expression
\be
\label{eq45}
E_1^\Omega(\Omega) = \left(1 + \frac{1}{3}\,j\,\frac{\md}{\md j}\right)H_1^\textrm{circ}(j),
\ee
which involves a crucial $j$ derivative [by contrast to the derivation of the function 
$H_1^\textrm{circ}(j)= [H^\textrm{tail ($s=r_{12}$)}]^\textrm{circ}(j)$ which involved no differentiation].

Applying Eq.\ \eqref{eq45} to Eq.\ \eqref{eq44} yields
\begin{align}
\label{eq46}
E_\textrm{$r_{12}$-tail}^\Omega(\Omega) &= -\frac{448}{15} \mu c^2 \nu x^5
\left[ \ln\left(\frac{4\me^\gE\Omega\, r_0^\textrm{circ}(j)}{c} \right)
\right.\nonumber\\[1ex]&\qquad\left.
-\frac{1}{7} \frac{\md\ln\Omega}{\md\ln j}
-\frac{1}{7} \frac{\md\ln r_0^\textrm{circ}(j)}{\md\ln j} \right],
\end{align}
where (on shell) $\ln\left(\frac{4\me^\gE\Omega\,r_0^\textrm{circ}(j)}{c} \right)=\ln(4\me^\gE x^{\frac12})$.
As $r_0^\textrm{circ}(j)=mj^2$, the contribution of the last term\footnote{
Note, in passing, that if we were considering the $s$-scaled tail
(with a fixed scale $s$) the latter term would be
\mbox{$-\frac{1}{7}\md\ln s/\md\ln j=0$}.} in the bracket is $-2/7$.
Moreover, as our Delaunay-based method led us to have $\widehat\Omega=\widehat\Omega_0^\textrm{circ}(j)=j^{-3}$ in Eq.\ \eqref{eq44}
[see notably Eq.\ (4.11) in Ref.\ \cite{Damour:2015isa}], the penultimate term in the bracket yields the additional term
\be
\label{eq47}
-\frac{1}{7} \frac{\md\ln\Omega_0^\textrm{circ}(j)}{\md\ln j} = + \frac{3}{7} \, ,
\ee
so that the bracket in Eq.\ \eqref{eq46} becomes
\be
\label{eq48}
\ln\left(\frac{4\me^\gE\Omega\,r_0}{c}\right) + \frac{3}{7} - \frac{2}{7}
= \ln\left(\frac{4\me^\gE\Omega\,r_0}{c}\right) + \frac{1}{7} \, .
\ee
By contrast, the result claimed in Ref.\ \cite{Bernard:2015njp} for that bracket
[see Eq.\ (5.30) there; note that in Eq.\ \eqref{eq46} we factored 448/15,
instead of the 224/15 they factor in their Eq.\ (5.30), to better exhibit the role of $\ln(\Omega r_0)$]
reads
\be
\label{eqb3fm}
\ln\left(\frac{4\me^\gE\Omega\,r_0}{c}\right)-\frac{2}{7} \, .
\ee
The shift $+\frac{3}{7}$, Eq.\ \eqref{eq47}, connecting the result \eqref{eqb3fm} of Ref.\ \cite{Bernard:2015njp}
to our \cite{Damour:2014jta,Damour:2015isa} result \eqref{eq48}
is the basis of the claim of Ref.\ \cite{Bernard:2015njp} that our value of the constant $C$
should be changed into $C+\frac{3}{7}$ [see Eq.\ (5.31) in Ref.\ \cite{Bernard:2015njp}].

We disagree with this conclusion of Ref.\ \cite{Bernard:2015njp}
because of the Delaunay-based logic, recalled above, that unescapably leads
to the result \eqref{eq48} of Refs.\ \cite{Damour:2014jta,Damour:2015isa}. 
In the next section, we explain why the treatment of Ref.\ \cite{Bernard:2015njp}
(which, at the effective level, consists in treating $\Omega$  as a constant,
instead of a function of $j$), though justified within a certain (nonlocal) framework,
cannot be correctly combined with the use of the Hamiltonian as conserved energy.

\section{Subtleties in the definition of the energy in the presence of nonlocal effects}

In this section we discuss subtle aspects of the order reduction of nonlocal actions
which are automatically incorporated within our approach,
but which can easily lead to apparent paradoxes and/or errors
when trying to define a conserved energy along circular orbits.

It will be convenient to use an even simpler toy model than the one used above to explain
these subtleties, and to further confirm the validity of our result \eqref{eq48}.

Let us start (as above) with a nonlocal interaction described (in 
Poincar\'e-Delaunay variables) by a toy action of the form
\be
S = \int \left[\Lambda\,\md\lambda + I_r\,\md\varpi - H^\textrm{nonloc}(t)\,\md t\right],
\ee
with
\begin{multline}
\label{eq49}
H^\textrm{nonloc}(t) = -\frac{1}{2\Lambda^2}
\\[1ex]
+ \varepsilon \int\md\tau \mu(\tau)\Lambda'^n\Lambda^n
\left[\cos(\lambda'-\lambda) + O(e)\right],
\end{multline}
where, as above, $\Lambda\equiv\Lambda(t)$,
$\Lambda'\equiv\Lambda(t')\equiv\Lambda(t+\tau)$, etc.

Above, we explained how to completely reduce the nonlocality of this action
[i.e.\ the dependence of $H^\textrm{nonloc}(t) $ on dynamical variables at time $t' \neq t$] to a purely local dependence.
In this reduction the term $ \int\md\tau \mu(\tau)\Lambda'^n\Lambda^n \cos(\lambda'-\lambda)$ was replaced
by $\int\md\tau \mu(\tau)\Lambda^{2n} \cos(\Omega(\Lambda) \tau)$ with $\Omega(\Lambda)=1/\Lambda^3$.
And we have then seen  the crucial role played by the $\Lambda$ dependence of $\Omega$
(corresponding to the $J$ dependence discussed in the previous section).

In order to better see the crucial transition from the original $\cos\big(\lambda(t')-\lambda(t)\big)$
(where there is no apparent $\Lambda$ dependence)
to the final $\cos\big(\Omega(\Lambda)(t'-t)\big)$,
we shall study a simple toy model where the original oscillatory factor
$\cos\big(\lambda(t')-\lambda(t)\big)\equiv\cos\Big(\frac{\lambda(t')-\lambda(t)}{t'-t}(t'-t)\Big)$
is replaced by $\cos\big(\dot\lambda\,(t'-t)\big)$ with $\dot\lambda\equiv\md\lambda/\md t$.

A motivation for considering such a simplified dynamics is as follows.
If one were formally considering a nonlocality
kernel $\mu(\tau)$ which decays sufficiently fast for large $|\tau |$'s 
[say $\mu(\tau)=\exp[-\frac{1}{2}(\frac{\tau}{\sigma})^2]/\sqrt{2\pi\sigma^2}$],
one could formally replace 
in any nonlocal Hamiltonian $H^{\rm nonloc}= \int \md \tau \mu(\tau) f(q, q', p, p')$,
the time-shifted variables $q'=q(t+\tau)$ and $p'=p(t+\tau)$ by their Taylor expansions,
$q'= q + \tau \dot q + \frac12 \tau^2 \ddot q + \cdots$, etc. This then replaces the original nonlocality 
by a generalized Ostrogradski Hamiltonian depending on the infinite tower of derivatives
of the phase-space variables:  $H^{\rm ostro}(q,p; \dot q, \dot p, \ddot q, \ddot p, \cdots)$. 
The theory of the order reduction of such {\it quasilocal}, Ostrogradski-type Hamiltonians
has been abundantly treated in the literature,
particularly in the context of the PN-expanded dynamics of binary systems
(see, e.g., Refs.\ \cite{Schafer:1984mr,Damour:1990jh,Damour:1985mt,Damour:1999cr}).
By considerations that would be too long to explain here, one can then see that for nonlocalities
of the Delaunay type $ \int\md\tau \mu(\tau)\Lambda'^n\Lambda^n \cos(\lambda'-\lambda)$,
in which one Taylor expands the time-shifted variables, e.g.\
$\lambda' = \lambda(t+\tau) = \lambda + \tau\dot{\lambda} + \frac{1}{2}\tau^2\ddot{\lambda} + \cdots$,
$\Lambda' = \Lambda(t+\tau) = \Lambda + \tau\dot{\Lambda} + \frac{1}{2}\tau^2\ddot{\Lambda} + \cdots$,
the crucially delicate nonlocality is only contained  in the term $\tau\dot{\lambda}$.
The other ones are essentially equivalent to double-zero terms [because of some special structure 
of the relevant Delaunay-like Hamiltonians, notably a $\mathbb{Z}_2$ symmetry transformation combining time reversal
($t\to-t$, $t'\to-t'$, $\tau\to-\tau$)
and angle reversals ($\lambda\to-\lambda$, $\varpi\to-\varpi$),
and keeping the action variables fixed].

Independently of this motivation, we shall simply use here, for pedagogical purposes,
as a simple model of the nonlocal action \eqref{eq49}, the following first-order Ostrogradski Hamiltonian:
\begin{multline}
\label{eq54}
H^\textrm{ostro}_\textrm{toy}(\lambda,\Lambda, \varpi, I_r;\dot{\lambda},\dot{\varpi})
= - \frac{1}{2\Lambda^2}
\\[1ex]
+ \varepsilon \int\md\tau\,\mu(\tau)\,\Lambda^{2n}
\left[\cos(\dot{\lambda} \, \tau) + O(e)\right],
\end{multline}
where the $\dot\varpi$ dependence only occurs within the $O(e)$ remainder.

Note that this first-order Ostrogradski Hamiltonian has a {\it quasilocal}
(but \emph{not} local) structure in that it depends not only on the phase-space variables $(q,p)$,
but also on some of their (first) derivatives (here a dependence on the two $\dot q$'s).
Though the ``nonlocality" associated with the presence of these derivatives  in the Hamiltonian \eqref{eq54} might seem trivial,
it will actually display the origin of the discrepancy between Ref.\ \cite{Bernard:2015njp} and us,
and show why the claims of Ref.\ \cite{Bernard:2015njp} are unfounded.

In Ref.\ \cite{Damour:1999cr} we have investigated in detail Ostrogradski Hamiltonians depending
on $\dot{q}$, $\dot{p}$ in addition to $q$, $p$.
Our main results were the following.
First, when considering a general $H^\textrm{ostro}(q,p;\dot{q},\dot{p})$
[with action $S_\textrm{ostro}=\int(p\dot{q}-H^\textrm{ostro})\md t$],
there exists a \emph{conserved energy} canonically associated with such an Ostrogradski Hamiltonian,
which is \emph{not} given by the Hamiltonian but rather by the following (Noether-theorem-deduced) quantity:
\begin{multline}
\label{eq55}
\mathcal{E}^\textrm{cons}(q,p;\dot{q},\dot{p}) = H^\textrm{ostro}(q,p;\dot{q},\dot{p})
\\[1ex]
- \dot{q}\frac{\partial H^\textrm{ostro}}{\partial\dot{q}}
- \dot{p}\frac{\partial H^\textrm{ostro}}{\partial\dot{p}}.
\end{multline}

Second, when considering the case of interest here where
\be
\label{eq56}
H^\textrm{ostro}(q,p;\dot{q},\dot{p})
= H_0(q,p) + \varepsilon H_1(q,p;\dot{q},\dot{p}),
\ee
one can reduce [modulo $O(\varepsilon^2)]$ $H^{\rm ostro}$
to an {\it ordinary} Hamiltonian $H^{\rm ordin}(q',p')$
[with action $S^{\rm ordin} = \int (p' \md q' - H^{\rm ordin}\md t)$]
by means of the following shifts:
\be
\label{eq57}
q' = q + \frac{\partial H^\textrm{ostro}}{\partial\dot{p}},
\quad
p' = p - \frac{\partial H^\textrm{ostro}}{\partial\dot{q}}.
\ee
In addition the {\it function} $H^{\rm ordin}(q,p)$
is obtained by the naive (``incorrect'') order reducing
of $H^{\rm ostro} (q,p ; \dot q , \dot p)$,
i.e.\ by inserting the lower-order equations of motion in $H^{\rm ostro}$,
\be
\label{eq58}
H^{\rm ordin}(q,p)
= H^\textrm{ostro}\left(q,p;\frac{\delta H_0}{\delta p},-\frac{\delta H_0}{\delta q}\right)
+ O(\varepsilon^2).
\ee
And, finally, the (Noether) conserved energy
(expressed in terms of the original variables)
is (numerically) equal to the (order-reduced) ordinary Hamiltonian $H^{\rm ordin}$
[expressed in terms of the corresponding {\it shifted} variables
$q'$, $p'$ of Eq.\ \eqref{eq57}]:
\be
\label{eq59}
\mathcal{E}^\textrm{cons}(q,p;\dot{q},\dot{p})
= H^{\rm ordin}(q',p') + O(\varepsilon^2).
\ee
Let us now apply the general results \eqref{eq55}--\eqref{eq59}
(recalled from Ref.\ \cite{Damour:1999cr}) to our toy model \eqref{eq54}.
We first note the existence of a (Noetherian) conserved energy, namely
\begin{multline}
\label{eq60}
\mathcal{E}^\textrm{cons}_\textrm{toy}(\lambda,\Lambda;\dot{\lambda},\dot{\varpi})
=  H^\textrm{ostro}_\textrm{toy}(\lambda,\Lambda;\dot{\lambda},\dot{\varpi})
\\[1ex]
- \dot{\lambda} \frac{\partial H^\textrm{ostro}_\textrm{toy}}{\partial\dot{\lambda}}
- \dot{\varpi} \frac{\partial H^\textrm{ostro}_\textrm{toy}}{\partial\dot{\varpi}}.
\end{multline}
In the circular limit the last term 
$- \dot{\varpi} {\partial H^\textrm{ostro}_\textrm{toy}}/{\partial\dot{\varpi}}$ vanishes
because of the vanishing eccentricity\footnote{
Here, one should use regular Poincar\'e variables to rigorously show
that the $O(e)$ terms in \eqref{eq54} do not contribute when $e \to 0$.}.
On the other hand the penultimate term {\it does not} vanish in the circular limit
because it involves a $\tau$-even integrand:
\be
\label{eq61}
\frac{\partial}{\partial\dot{\lambda}}
\int\md\tau\,\mu(\tau)\cos(\dot{\lambda}\tau)
= - \int\md\tau\,\mu(\tau)\,\tau\sin(\dot{\lambda}\tau),
\ee
so that
\begin{multline}
\label{eq62}
\mathcal{E}^\textrm{cons}_\textrm{circ}(\Lambda,\dot{\lambda})
= -\frac{1}{2\Lambda^2}
\\[1ex]
+ \varepsilon \Lambda^{2n} \int\md\tau\,\mu(\tau)
\left[\cos(\dot{\lambda}\tau) + \dot{\lambda}\tau\sin(\dot{\lambda}\tau)\right].
\end{multline}
In the circular limit the extra term $-\dot\lambda\,\partial H / \partial \dot\lambda$
is independent of time and cannot therefore be guessed simply by looking
at the equations of motion. 

In other words, our toy model shows that, when working with a non-order-reduced, nonlocal
Hamiltonian, the Noetherian conserved energy differs from the Hamiltonian, even in
the circular limit. This is another way to view (part of) the discrepancy with the derivation
of the function $E(\Omega)$ in Ref.  \cite{Bernard:2015njp}. Indeed, the latter reference
works within a  (non-order-reduced) nonlocal framework, but uses as conserved circular energy 
the circular-reduced Hamiltonian itself,
i.e., in our toy model, simply $H_{\rm toy}^{\rm ostro} (\Lambda , \dot\lambda)$,
which differs from the ``good'' (Noetherian) conserved energy
\eqref{eq60} and \eqref{eq62}.

Another useful outcome of the results \eqref{eq55}--\eqref{eq59} recalled above is that,
while there is no need to shift the angular variables $\lambda , \varpi$
(because $\delta\lambda= \partial H^{\rm ostro} / \partial \dot\Lambda = 0$,
$\delta\varpi=\partial H^{\rm ostro} / \partial \dot I_r = 0$),
one must shift the action variables if one wishes to have an order-reduced dynamics.
The first fact says in particular that, in the circular limit,
the orbital angular frequency $\dot\lambda = \dot \lambda'$ is unambiguous
and is given either by the (functional) derivative
\begin{align}
\label{eq63}
\dot\lambda &= \frac{\delta H^\textrm{ostro}_\textrm{toy}(\Lambda,\dot{\lambda})}
{\delta\Lambda}
\nonumber\\[1ex]
&= \frac{1}{\Lambda^3} + 2n\varepsilon\Lambda^{2n-1}
\int\md\tau\,\mu(\tau)\cos(\dot{\lambda}\tau),
\end{align}
in which the frequency $\dot \lambda$ in the argument of the cosine
is treated as a constant, or [up to $O(\varepsilon^2)$ corrections]
by the (ordinary) derivative of the order-reduced, ordinary Hamiltonian
\begin{multline}
\label{eq64}
H^\textrm{ordin}_\textrm{toy}(\lambda',\varpi',\Lambda',I_r')
= -\frac{1}{2\Lambda'^2}
\\[1ex]
+ \varepsilon \Lambda'^{2n} \int\md\tau\,\mu(\tau)
\left[\cos\left(\frac{\tau}{\Lambda'^3}\right) + O(e)\right],
\end{multline}
which yields (in the circular limit) an orbital frequency given by
\begin{align}
\label{eq65}
\dot{\lambda}' &= \frac{\partial H^\textrm{ordin}_\textrm{toy}}{\partial\Lambda'}
\nonumber\\[1ex]
&= \frac{1}{\Lambda'^3} + \varepsilon \frac{\partial}{\partial\Lambda'}
\left[ \Lambda'^{2n} \int\md\tau\,\mu(\tau)
\cos\left(\frac{\tau}{\Lambda'^3}\right)\right].
\end{align}

Contrary to Eq.\ \eqref{eq63} the $\Lambda'$ derivative on the rhs of Eq.\ \eqref{eq65}
now acts both on the prefactor $\Lambda'^{2n}$ and on the argument of the cosine.
Note that in the real 4PN problem the power $2n$ is equal to $-10$
and the weight function $\mu(\tau)$ is either ${\rm Pf}_{2s/c} \vert \tau \vert^{-1}$
or ${\rm Pf}_{2r_{12}/c} \vert \tau \vert^{-1}$ (as one chooses).
Using, say, the latter ($r_{12}$-) choice, and, e.g., Eq.\ (5.8) of Ref.\ \cite{Damour:2014jta}, one has then
\be
\label{eq66}
\varepsilon\Lambda^{-10}\int_{-\infty}^{+\infty}\md\tau\,\mu(\tau)\cos(\dot{\lambda}\tau)
= -\varepsilon\Lambda^{-10}\ln\left(\frac{4 \me^{\gE}\dot{\lambda}r_{12}}{c}\right) \, .
\ee
The result \eqref{eq66} displays the structure of the tail contribution to the Hamiltonian,
as in Eq. \eqref{eq44}. However, we now see that there are two different ways of dealing with this contribution.
(The consistency of these two different ways will be discussed below.)
The difference between the two calculations \eqref{eq63} and \eqref{eq65} parallels what we explained above
about the difference between the prescription used by Ref.\ \cite{Bernard:2015njp}
(treating $\dot\lambda$ as a constant within the argument of the tail logarithm)
and the prescription used by us \cite{Damour:2014jta,Damour:2015isa} (where $\Omega$ was a function of $j$).
Indeed, there are two different methods.
The first method is to stay within a non-order-reduced, nonlocal framework (as is the case of Ref.\ \cite{Bernard:2015njp}),
and then consider the orbital frequency $\dot{\lambda}$ as being \textit{a priori} unrelated to $j$
[so that the additional term \eqref{eq47} is \textit{a priori} absent as in the calculation of Ref.\ \cite{Bernard:2015njp}],
but, one must then take as conserved energy the quantity \eqref{eq60}, which differs from the Hamiltonian in the circular limit
(contrary to what was used in Ref.\ \cite{Bernard:2015njp}).
The second method is to work within an order-reduced,
local Hamiltonian (as is the case of Ref.\ \cite{Damour:2015isa}), in which case one must
take into account the additional term \eqref{eq47}, and one can (and must) use
as conserved energy the (order-reduced, ordinary) Hamiltonian when computing
the function $E(\Omega)$ (as was done in Refs.\ \cite{Damour:2014jta,Damour:2015isa}).

The detailed reasoning we just made shows that the calculation \eqref{eq65}
[involving the ordinary, order-reduced Hamiltonian \eqref{eq64}]
is not only self-consistent (as we already argued above) but, most importantly,
is consistent with the {\it nonlocal} dynamics
of the original Ostrogradski Hamiltonian \eqref{eq54}, under the following two conditions:
(i) the conserved energy associated with the nonlocal description \eqref{eq54}
should not be equated (contrary to the prescription used in Ref.\ \cite{Bernard:2015njp}),
when considering the circular limit, with the value of $H^{\rm ostro}$,
but rather with Eqs.\ \eqref{eq60} and \eqref{eq62},
and (ii) the variables $\lambda, \Lambda, \varpi, I_r$
entering the (weakly) nonlocal dynamics \eqref{eq54}
cannot be equated with the variables $\lambda' , \Lambda' , \varpi' , I'_r$
entering the corresponding order-reduced ordinary dynamics \eqref{eq64}.
More precisely, the link between the two sets of variables
is $\lambda'=\lambda$, $\varpi' = \varpi$ but
\begin{subequations}
\begin{align}
\label{eq66a}
\Lambda' &= \Lambda - \frac{\partial H^\textrm{ostro}_\textrm{toy}}{\partial\dot{\lambda}}
= \Lambda + \varepsilon \Lambda'^{2n} \int\md\tau\,\mu(\tau)\tau\sin(\dot{\lambda} \tau),
\\[1ex]
\label{eq66b}
I_r' &= I_r - \frac{\partial H^\textrm{ostro}_\textrm{toy}}{\partial\dot{\varpi}}
= I_r + O(e).
\end{align}
\end{subequations}
It is easily checked that the shift \eqref{eq66a} ensures the consistency
between the two (different) functions $\Lambda \to \dot\lambda$,  Eq.\ \eqref{eq63},
and $\Lambda' \to \dot\lambda' = \dot\lambda$, Eq.\ \eqref{eq65}.
It is moreover checked that the conserved energy
${\mathcal E}_{\rm circ}^{\rm cons} (\Lambda , \dot\lambda)$, Eq.\ \eqref{eq62},
becomes numerically equal to the ordinary Hamiltonian energy \eqref{eq64}
when also taking into account the shift \eqref{eq66a}.

Summarizing so far: we have explicitly shown, by two different approaches
(one of them being the one used in Refs.\ \cite{Damour:2014jta,Damour:2015isa},
and the other being as close as possible to the one used in Ref.\ \cite{Bernard:2015njp}),
that the correct $(r_{12}$-scaled) tail contribution to the (Noetherian) conserved energy
(along circular orbits) expressed in terms of the orbital frequency
is given by Eqs.\ \eqref{eq46} and \eqref{eq48},
in agreement with our results \cite{Damour:2014jta,Damour:2015isa}.

\section{Incompatibility between the $C$-modified Fokker-action result of Ref.\ \cite{Bernard:2015njp} and self-force results}

An immediate consequence of the result of the previous section is
that the determination in Ref.\ \cite{Bernard:2015njp} of the ratio
between the IR cutoff scale $r_0$ and the scale $s_0 = r'_0 \me^{-11/12}$
entering the computation of the tail-transported near-zone metric perturbation \cite{Blanchet:1987wq}
is incorrect. In other words, in terms of the notation of Ref.\ \cite{Damour:2014jta}
[and of Eqs.\ \eqref{eq4}--\eqref{eq6} above]
one should add to the Hamiltonian of Ref.\ \cite{Bernard:2015njp}
an extra term $\Delta C \, F[\vecx_a,\vecp_a]$ with
\be
\label{eq67}
\Delta C = -\frac37 .
\ee
If we understand correctly the notation used in Ref.\ \cite{Bernard:2015njp},
their (ambiguity) parameter $\alpha$ [in their Eq.\ (3.9)]
is equivalent to our (ambiguity) parameter $-C$ modulo a fixed, additive constant
linked to the difference between the two schemes.
It would then mean that the correct value of their $\alpha$ should be
\be
\label{eq68}
\alpha^{\rm new}_{\rm B^3FM} = \alpha_{\rm B^3FM} + \frac{3}{7}
= \frac{811}{672} + \frac{3}{7} = \frac{157}{96},
\ee
instead of the value, here denoted $\alpha_{\rm B^3FM}$, in their Eq.~(4.30).
(Note that the denominator of $\alpha^{\rm new}_{\rm B^3FM}$ is simpler
than that of $\alpha_{\rm B^3FM}$: namely $96 = 2^5 \times 3$
while $672 = 7 \times 96 = 2^5 \times 3 \times 7$.)

Modifying the Hamiltonian of Ref.\ \cite{Bernard:2015njp} by $+\Delta C \, F$
then modifies the difference $H^{\rm new}_{\rm B^3FM} - H_{\rm DJS}$
into an expression of the same form as the rhs of Eq.\ \eqref{eq7}
but with the following new values of the three coefficients $(a,b,c)$:
\be
\label{eq69}
(a,b,c)^{\rm new}_{\rm B^3FM} = (a,b,c)_{\rm B^3FM}
+ \Delta C\,\frac{16}{15}\,(-11,12,0).
\ee
Inserting $\Delta C = -\frac37$ from Eq.\ \eqref{eq67},
and using Eq.\ \eqref{eq8} above, yields
\be
\label{eq70}
(a,b,c)^{\rm new}_{\rm B^3FM} = \frac{1}{315}\,(3013,-902,902).
\ee
The latter result is equivalent
[modulo a gauge transformation, Eq.\ \eqref{eq11}, with $g = 288/315$]
to the rhs of Eq.\ (5.34b) in Ref.\ \cite{Bernard:2015njp},
but with the different meaning that
it is the Hamiltonian of Ref.\ \cite{Bernard:2015njp} on the lhs
which has to be shifted instead of $H_{\rm DJS}$.

Inserting Eq.\ \eqref{eq70} into our general formulas \eqref{eq9}--\eqref{eq10}
then leads to the following additional contributions
(compared to our results \cite{Damour:2015isa})
to the EOB potentials $A$ and $\bar{D}$
entailed by the above modified version of the Hamiltonian of Ref.\ \cite{Bernard:2015njp}:
\begin{align}
\label{eq71}
\delta^\textrm{new} A &= 0,
\\[1ex]
\label{eq72}
\delta^\textrm{new}  \bar{D} &= -\frac{34}{9}\,\nu\,u^4.
\end{align}

The first result, Eq.\ \eqref{eq71}, shows that our prescription \eqref{eq67}
for modifying the result of Ref.\ \cite{Bernard:2015njp} now yields a value of the EOB $A$
potential at 4PN in agreement with our result \cite{Damour:2014jta,Damour:2015isa}
(and with the many existing SF computations of $A$ recalled above).
On the other hand, we see in Eq.\ \eqref{eq72} that, even after our correction,
the 4PN contribution to the EOB $\bar{D}$ potential predicted by the Fokker-action,
harmonic-gauge computation of Ref.\ \cite{Bernard:2015njp} significantly differs
from our result \cite{Damour:2015isa}. 

In terms of the gauge-invariant 4PN-level parameters $a_5^c$, $\bar d_4^c$, $\rho_4^c$ introduced above, the results 
\eqref{deltaa5}, \eqref{deltad4}, and \eqref{deltarho4} above are now replaced by
\begin{align}
\label{deltaa5'}
\delta^{\textrm{B$^3$FM}'} a_5^c &= 0 ,
\\[1ex]
\label{deltad4'}
\delta^{\textrm{B$^3$FM}'} \bar{d}_4^c &=  -\frac{34}{9}= -3.77777\cdots ,
\\[1ex]
\label{deltarho4'}
\delta^{\textrm{B$^3$FM}'} {\rho}_4^c &=  -\frac{34}{9}= -3.77777\cdots  \, .
\end{align}

As we already said above, there are many
(analytical and numerical) SF results showing that the 4PN coefficients $\bar{d}_4^c$ or $\rho_4^c$ 
nicely agree with the ADM result \cite{Damour:2015isa}.
(Let us, in particular, recall the recent {\it analytical} confirmations \cite{Bini:2015bla,Hopper:2015icj}.)
Numerical tests making use of the eccentric first law \cite{Tiec:2015cxa}
yield a  limit $\vert \delta \bar{d}_4^c \vert^{\rm num} < 0.05$ on 
any possible deviation $\delta \bar{d}_4^c$ away from our analytical result
(see discussion and references, in Ref.\ \cite{Bini:2015bla} and above).
As for direct dynamical limits on the 4PN-level coefficient of the precession function $\rho(u)$
we have displayed them in Eqs.\ \eqref{drho4barack} and \eqref{drho4meent} above.
While the old limit \eqref{drho4barack} \cite{Barack:2010ny} would be compatible with the modified result \eqref{deltarho4'},
the recent limit  \eqref{drho4meent} \cite{vandeMeent:2016} is in violent disagreement with Eq.\ \eqref{deltarho4'}.

\section{Suggestion for adding more IR ambiguity parameters in Ref.\ \cite{Bernard:2015njp}}

In view of the strong discrepancy \eqref{deltad4'} and \eqref{deltarho4'},
remaining after making use of the sole IR ambiguity parameter $C=-\alpha+\textrm{const}$,
we suspect that the harmonic Fokker action derived in Ref.\ \cite{Bernard:2015njp}
has to be modified not only by the shift \eqref{eq67} of $C$ (or $-\alpha$),
but by suitable shifts of further {\it ambiguity parameters}
entering their present computation.

An independent argument suggesting the presence of more ambiguities
in their calculations than the mere $C\,F[\vecr,\vecp]$ ambiguity
(which is also present in our ADM calculation,
and was only fixed by using the analytical result of Ref.\ \cite{Bini:2013zaa}) is presented next.

On the one hand, though the ADM calculation is not manifestly Poincar\'e invariant,
and involves both UV and IR divergences that we had to regularize, our final
result is Poincar\'e invariant as is, without having to invoke any correcting contact transformation.
In particular, the 4PN ADM Hamiltonian is {\it manifestly} invariant under spatial translations,
i.e.\ only depends on the relative position ${\mathbf{y}}_{12}={\mathbf{y}}_1-{\mathbf{y}}_2$
of the two particles.

On the other hand, the core of the harmonic-coordinates calculation of Ref.\ \cite{Bernard:2015njp}
is the evaluation of the (local) Fokker Lagrangian [their Eqs.\ (2.20) and (4.20)]
\be
\label{Lg}
L_g = {\rm FP} \int\md^3x\, \left(\frac{r}{r_0}\right)^B\mathcal{L}_g,
\ee
where
\be
\label{eq73}
\mathcal{L}_g
\sim c^4\,
\big[\bar{h}\,\Box\bar{h} + \bar{h}\,\partial\bar{h}\,\partial\bar{h}
+ \cdots
+ \bar{h}\,\bar{h}\,\bar{h}\,\bar{h}\,\partial\bar{h}\,\partial\bar{h} \big],
\ee
in which one inserts the formal, near-zone PN expansion of the metric variables,
computed in harmonic coordinates.
Both the harmonic Fokker Lagrangian density ${\mathcal L}_g$
and the (time-symmetric) harmonic metric are formally Poincar\'e covariant.
The Lagrangian $L_g$ should then vary, under Poincar\'e transformations,
as the time component of a four-vector, except if the (non-Poincar\'e-invariant)
IR regulator $ {\rm FP} (| {\bf x} |/r_0)^B $ introduces violations of Poincar\'e invariance.
Note, in particular, that all the ingredients of the calculation are {\it manifestly} invariant
under spatial translations, except for the IR regulator.
However, Bernard \textit{et al.\@} \cite{Bernard:2015njp} stated that
the raw result of the computation \eqref{eq73} is {\it not} translation invariant,
but depends on the individual positions ${\mathbf{y}}_A$ of the particles.
They found that the offending terms can be removed by some shifts $\mathbf{\xi}_A$
of the positions ${\mathbf{y}}_A$ so as to yield a manifestly Poincar\'e covariant
(modulo a time derivative) Lagrangian, say $L_g^{\rm shifted}$.
An inspection of the needed shifts (in their Appendix C) reveals
a very large number of offending non-translation-invariant terms
which are mostly connected with IR divergences [$\propto\ln(r_{12}/r_0)$].
Moreover, there are terms in the shifts $\mathbf{\xi}_A$
which combine both IR and UV divergences [$\propto\ln(r_{12}/r'_1)$].
By contrast with the ADM calculation which, in spite of  its nonmanifest Poincar\'e
invariance, ends up with a Poincar\'e-invariant (and, in particular, translation-invariant)
regularized dynamics, we consider that several features of the harmonic-coordinate
calculation strongly suggest the presence of more ambiguities in the final result for $L_g$
than the $CF[\vecr,\vecp]$ one, namely:
1) the large zoology of nontranslationally invariant IR-divergent terms;
2) the mixing of IR and UV divergences,
together with the use of a method which uses Hadamard regularization
as an intermediate step for UV divergences
(before ``correcting'' it by extra dimensional-regularization terms); and,
3) the known fact that harmonic coordinates imply a worse behavior in the outer near zone
than ADM coordinates [notably because the ten metric components $g_{\mu \nu}$ propagate with
the velocity of light in harmonic coordinates, while, in ADM coordinates,
only $h_{ij}^{\rm TT} = O(1/c^4)$ does so].

We therefore suggest that Bernard \textit{et al.\@} \cite{Bernard:2015njp} should
(similarly to what happened at the 3PN level when using only Hadamard-type regularization
\cite{Jaranowski:1997ky,Damour:1999cr,Blanchet:2000ub,Blanchet:2001aw})
acknowledge the presence of further ambiguities in their result,
parametrized by new ambiguity parameters.
We shall assume here that these ambiguities affect
the three discrepant terms \eqref{eq7} and only them.

In view of the unavoidable presence of the $CF[\vecr,\vecp]$ ambiguity,
and of the possibility of making the gauge transformation \eqref{eq11},
the minimal number of extra ambiguity parameters that need to be introduced is {\it one}.
For instance, we could introduce a single ambiguity
parametrized by adding to the result of Ref.\ \cite{Bernard:2015njp}
a term in the Hamiltonian of the form
\be
\label{eq74}
\delta^{\Delta a}H = +\Delta a \frac{G^4 M m_1^2 m_2^2}{c^8 r_{12}^4}
\bigg(\frac{\npi}{m_1}-\frac{\npii}{m_2}\bigg)^2,
\ee
which has the effect of changing $(a,b,c)$ into $(a + \Delta a , b , c)$.
Such a change does not affect the EOB $A$ potential
[which is $\propto2(b+c)$, Eq.\ \eqref{eq9}],
but affects the EOB $\bar{D}$ potential
[which is $\propto 2 (a + 4b)$, Eq.\ \eqref{eq10}].
We conclude that one must add to their result in addition to $\Delta CF$,
with $\Delta C = -3/7$, Eq.\ \eqref{eq7},
the extra contribution \eqref{eq74} with
\be
\label{eq75}
\Delta a = +\frac{17}9.
\ee

It is also possible that part of the ambiguities in their calculation
are parametrized by a transformation of the type \eqref{eq11},
but such a transformation has no gauge-invariant impact on the dynamics
as it is induced by a canonical transformation with generating function
$\propto\nu\,{\mathbf{n}}_{12}\cdot(\vecp_1 / m_1 - \vecp_2 / m_2)\,r_{12}^{-3}$
(corresponding to changes of $\vecr_1$ and $\vecr_2$ in a purely radial direction).
(Note, however, that even such ``gauge ambiguities''
will affect the computation of the radiative quadrupole moment of the system, see below.)

\section{Conclusions}

Let us summarize our main conclusions concerning the conservative
dynamics of a binary system at the 4PN approximation.

\begin{enumerate}

\item[$\bullet$] We have reviewed the numerous confirmations of the
correctness of the 4PN ADM dynamics \cite{Damour:2014jta,Damour:2015isa,Jaranowski:2015lha}.
When decomposed in powers of the symmetric mass ratio $\nu$, all its elements have been confirmed, piecewise,
by independent computations. In particular, the crucial linear-in-$\nu$ terms (at the first-self-force
level) have been confirmed, both analytically and numerically, by several self-force results.

\item[$\bullet$] By contrast we have computed the periastron precession induced by the recently
reported harmonic Fokker-action 4PN dynamics \cite{Bernard:2015njp} and found it to be in violent contradiction 
with  existent direct numerical self-force computations \cite{vandeMeent:2016}:
compare Eq.\ \eqref{deltarho4} to Eq.\ \eqref{drho4meent}.

\item[$\bullet$] After reviewing the logical basis and the consistency of the ADM-Delaunay-EOB derivation 
of Refs.\ \cite{Damour:2014jta,Damour:2015isa}, we pointed out two different flaws in the Fokker-action harmonic-coordinates
computation of Ref.\ \cite{Bernard:2015njp}: (i) their computation of the functional link,
along circular orbits, between the energy and the orbital frequency involves (when viewed within the approach of Sec.\ V)
a correct treatment of nonlocal (tail) effects
but a flawed assumption about the value of the physically relevant conserved energy;
(ii) after correcting for the previous issue,
there remains a violent incompatibility between their theoretically predicted periastron precession and recent dynamical self-force
computations [compare Eq.\ \eqref{deltarho4'} to Eq.\ \eqref{drho4meent}].

\item[$\bullet$] We suggest that the result of Ref.\ \cite{Bernard:2015njp} must be corrected by adding (at least) two
infrared ambiguity parameters  to the Hamiltonian form of their result: 
1) $\Delta C = - \Delta \alpha = -\frac37$, Eq.\ \eqref{eq67}; and 2) $\Delta a= + \frac{17}{9}$, 
Eqs.\ \eqref{eq74} and \eqref{eq75} [modulo a gauge transformation, Eq.\ \eqref{eq11}].
Let us again emphasize that our suggested corrections
represent only two rather minor adjustments among hundreds of terms that agree between 
two very difficult independent calculations, which have used different methods and different gauges.

\item[$\bullet$] Before attempting  a 4PN-level computation
of gravitational-wave emission (in the more convenient harmonic coordinates), 
it will be necessary to confirm,
within the framework of Ref.\ \cite{Bernard:2015njp}, the presence, and value,
of the couple of ambiguities (notably $\Delta a$ and $g$) mentioned above.

\end{enumerate}

\begin{acknowledgments}

T.D.\ thanks Leor Barack and Maarten van de Meent for informative email exchanges.
The work of P.J.\ was supported in part by the Polish NCN Grant
No.\ UMO-2014/14/M/ST9/00707.

\end{acknowledgments}

\appendix

\section{Fokker-type actions and tail Hamiltonian in the ADM formalism}

Before discussing several technical aspects of the (4PN-level) tail contribution to the ADM dynamics,
let us start by recalling some general features, common to the ADM action, and the harmonic-Fokker one.

Both actions are Fokker-type actions, obtained (as was done for the electromagnetic interaction in Ref.\ \cite{Fokker:1929})
by eliminating (or ``integrating out") the field, say $\phi$, in the total action, say $S_{\rm tot}[\phi, x_a]$, so as to obtain
a reduced action $S_{\rm red}[x_a]$ for the {\it conservative} dynamics of the particle worldlines $x_a^{\mu}(s_a)$. 
At the lowest order,
where one considers the linear coupling of the field to the worldline-distributed source $J[x_a]$, the elimination of the field
is obtained by a (formal) integration by parts leading to a replacement of the type (we omit indices, arguments and integration measures)
\be \label{nA1}
\int \left[ - \frac12 (\partial \phi)^2 + \phi J \right] = \int \frac12 J \, G \, J \, ,
\ee
where $G$ denotes the appropriate {\it time-symmetric}
(half-retarded-half-advanced) Green's function, which solves $ \Box \phi = - J$ as $\phi= G J$.
When considering, higher-order (nonlinear) couplings, one uses an iterative
solution of the field equations, and the elimination of the field then leads
to  a Feynman-like expression for the reduced action involving the concatenation
of sources, propagators and vertices, say
\be\label{nA2}
S_{\rm red}[x_a]= \int \frac12 J \, G \, J  + \int {\cal V}_3(G \, J, G \, J, G \, J) + \cdots,
\ee
where ${\cal V}_3$ denotes a cubic vertex.  See, e.g., Ref.\ \cite{Damour:1995kt} which
discussed in detail the corresponding diagrammatic representation of the reduced harmonic-gauge action.
See also Fig.\ 1 in Ref.\ \cite{Damour:2001bu}, and the discussion  below Eq.\ (2.9) there, for a diagrammatic
representation of the reduced ADM action.

Let us  note in passing that we disagree on some formal aspects
of the definition of a Fokker action in Ref.\ \cite{Bernard:2015njp}.
Indeed, contrary to the 4PN ADM computation which explicitly
uses the time-symmetric propagation of the TT degrees of freedom
(see Sec.\ IV of Ref.\ \cite{Jaranowski:2015lha}),
Ref.\ \cite{Bernard:2015njp} described their calculation
(at least for the quadrupolar tail contribution, but possibly not for
the main contributions to the local action) as if they formally worked with the nonconservative dynamics
of a binary system interacting via the {\it retarded} Green's function.
They then argue that the terms associated with radiation damping only contribute
time-derivative terms to the action they computed.
We understand technically why this works out formally;
however, we object to such a conceptual framework, because the use of a retarded Green's function physically implies a nonconservative dynamics.
Indeed, let us recall that, before formally eliminating the field $\phi$,
the variation of the original action $S_{\rm tot}[\phi, x_a]$
with respect to the particle worldlines $x_a^{\mu}$ yields (in general relativity)
the equations of motion of the particles (as geodesics in the metric $g_{\mu \nu}$).
This shows that when solving the field $\phi \sim g_{\mu \nu}$ by means of a retarded Green's function,
one necessarily gets a dynamics involving, retarded (and therefore dissipative \cite{Damour:1983tz}) interactions.
It is therefore conceptually incorrect to compute an action supposed to describe a conservative dynamics by using a retarded Green's function.
(From the technical point of view, the inconsistency of using a retarded field when
formally integrating by parts $S_{\rm tot}[\phi, x_a]$, while replacing $\phi$ as a functional of the worldlines,
shows up in nonvanishing surface terms.)
We therefore  define the conservative dynamics of gravitationally
interacting particles by integrating out the field obtained by iteratively solving the field equations by
time-symmetric propagators. It is not clear to us that, when working as we do here with nonlinear
aspects of general relativity, there is an alternative route for defining a conservative dynamics.
[The additional, nonlocal, time-antisymmetric radiation-reaction force needed to transform,
at the 4PN level, the conservative dynamics into the physically more directly meaningful dissipative
(causal) dynamics entailed by retarded interactions was exhibited in Sec.\ VI of Ref.\ \cite{Damour:2014jta}.]

Another point we want to make here is that, when working with such reduced actions, one can
consider that the dynamics is {\it defined} by the final worldline action \eqref{nA2}, and one
does not need to worry about the convergence properties of the initial field contributions
to $S_{\rm tot}[\phi,x_a]$ [such as $-\frac{1}{2}\int(\partial\phi)^2$], which are no longer
relevant after the elimination of the field $\phi$. For instance, in the trivial case of the linear
coupling \eqref{nA1}, the transition between  the initial total action $S_{\rm tot}[\phi,x_a]$,
and the reduced action $\int\frac12 J\,G\,J\,$ is well defined\footnote{
After factoring the usual infinity $-\int\md t\,E_\textrm{tot}$
associated with the dynamics of the center of mass of the system.}
for unbound (hyperbolic-type) motions.
On the other hand, if one considers instead  bound (elliptic-like) motions,
the spatial integral of the field energy (when considering the Hamiltonian) associated
with $\frac12 (\partial \phi)^2 $ is {\it linearly IR divergent} for bound (quasiperiodic) motions,
because both $\phi$ and its partial derivatives only decay as $O(1/r)$ at spatial infinity,
when considering, as must be done in a Fokker-action calculation, the time-symmetric solution $\phi_{\rm sym}$
[which is a superposition of standing waves of the rough type $\cos (\omega r)/r$].
(Related IR divergences have been recently pointed out in Ref.\ \cite{Pound:2015wva},
within the context of second-order gravitational self-force theory.)
But, we can consider that the  dynamics of bound motions is {\it defined}
by taking the analytic continuation (from the unbound case to the bound one) of
the (\textit{a priori} better IR-behaved) reduced action $S_{\rm red}[x_a]$, Eq.\ \eqref{nA2}.
[For instance, the linear interaction $\int \frac12 J \, G \, J \, $
features no IR divergences linked to the slow spatial decay of the (eliminated) field $\phi$.]
We adopt here this attitude (for the reduced ADM action).
This is why we wished to point out in Sec.\ IV that the oscillatory nature
of the integrand $I'^{(3)}_{ij} I_{ij}^{(3)}$ in the nonlocal piece of the 4PN action
ensured the  convergence of the integral over the relative time $\tau=t'-t$.
We leave to future work a more detailed analysis of the convergence properties of the ADM-Fokker action.

Reference \cite{Damour:2014jta} showed the necessity of including in the ADM action a
tail-related contribution $-\int\md t H^\textrm{tail sym\,($s$)}$, with
\begin{align}
\label{eqA1}
H_{\rm 4PN}^\textrm{tail sym\,($s$)}(t)
&= -\frac{1}{5}\frac{G^2 M}{c^8} \,I_{ij}^{(3)} (t)
\nonumber\\[1ex]
&\quad \times\,\Pf_{2s/c} \int_{-\infty}^{+\infty}
\frac{\md v}{\vert v \vert} \, I_{ij}^{(3)} (t+v),
\end{align}
by recalling the existence (first shown by Blanchet and Damour \cite{Blanchet:1987wq}) of a fundamental breakdown of the standard PN scheme at the 4PN level.
Indeed, when computing \emph{conservative} effects,
as is appropriate both in the computation of the reduced ADM action,
and in that of the harmonic action, the standard PN scheme is based 
on a formal near-zone expansion of the time-symmetric gravitational propagator, say
\begin{align}
\label{eqA2}
\Box^{-1}_\textrm{sym} &= \left(\Delta - \frac1{c^2}\,\partial_t^2\right)^{-1}
\nonumber\\[1ex]
&= \left(\Delta^{-1} + \frac1{c^2}\Delta^{-2}\partial_t^2 + \cdots\right) \delta(t-t').
\end{align}
At the 4PN level the \emph{near-zone} solution
generated by using the PN expansion \eqref{eqA2} is incomplete (as a near-zone solution),
and must be completed by adding to it a 4PN-level \emph{homogeneous} solution
of the (linearized) field equations (regular within the source)
which is a \emph{nonlocal-in-time}  functional
of the dynamical configuration of the system (related to tail effects).

Reference \cite{Blanchet:1987wq} showed that in a suitable coordinate gauge
the additional, 4PN-level, tail-related homogeneous metric perturbation
$h^\textrm{nonloc}_{\mu\nu\,\textrm{4PN}}$ contains only
a quadrupolar time-time component of the form
\be
\label{eqA3}
h^\textrm{nonloc}_{00\,\textrm{4PN}} = \frac{1}{2} x^i x^j H^\textrm{nonloc}_{ij}(t),
\ee
with the nonlocal function of time $H^\textrm{nonloc}_{ij}(t)$ given by
\be
\label{eqA4}
H^\textrm{nonloc}_{ij}(t) = -\frac{8}{5}\frac{G^2M}{c^{10}}
\Pf_{2s/c} \int_{-\infty}^{+\infty}
\frac{\md\tau}{\vert\tau\vert} \, I_{ij}^{(6)}(t+\tau).
\ee

The pure, quasi-Newtonian, nature of the additional (nonlocal)
metric perturbation \eqref{eqA3} allowed us, in Ref.\ \cite{Damour:2014jta},
by considering the corresponding additional term in the equations of motion of the system,
to (i) derive the corresponding additional (time-symmetric) nonlocal contribution
\eqref{eqA1} to the action (modulo a time derivative)
and, (ii) to check the consistency with the purely \emph{retarded} nature of the nonlocality
in the equations of motion when adding the (time-antisymmetric)
radiation-reaction force $\mathcal{F}^\textrm{rad reac}_\textrm{4PN}$.

Reference \cite{Bernard:2015njp} formally showed how to recover the nonlocal contribution
\eqref{eqA1} from the harmonic-gauge Fokker action, without considering the equations of motion,
and without using the simplifying features of the special gauge leading to Eq.\ \eqref{eqA3} only.
We wish here both to show how the same result can be achieved within the ADM formalism
and why the procedure advocated in Ref.\ \cite{Bernard:2015njp} is, in our opinion, exceedingly formal and actually
ill-defined because of its sensitivity to nonconvergent integrations by parts.

In the ADM formalism, before eliminating the TT variables, the Routhian\footnote{
The Routh functional is a Hamilton function for the bodies but a Lagrange functional
for the field degrees of freedom.}
\cite{Jaranowski:1997ky} reads (in the following we often use without warning
the ADM convention $16\pi G=1$)
\begin{multline}
\label{eqA5}
R[\vecx_a,\vecp_a,{\hTT ij},{\hTTdot ij}]
= \int\md^3x \left[ \frac{1}{4}(\partial_k{\hTT ij})^2
\right.\\[1ex]\left.
- \frac{1}{4c^2}(\hTTdot ij)^2
- \frac{1}{4}{\hTT ij}J_{ij} + \cdots \right],
\end{multline}
where the leading-order source term for ${\hTT ij}$ reads
\be
\label{eqA6}
J_{ij} = 2\sum_a\frac{p_{ai}p_{aj}}{m_a}\delta(\vecx-\vecx_a)
+ \frac{1}{2} \partial_i\phi\, \partial_j\phi + \cdots,
\ee
and where the ellipsis denotes higher-order terms that we shall not explicitly need here.
Note that the longitudinal field degrees of freedom, $\phi$ and ${\pit ij}$,
are viewed as being expressed in terms of the other degrees of freedom
($\vecx_a,\vecp_a,{\hTT ij},{\hTTdot ij}$) by using the constraints
(which are elliptic in nature).

The key point is that the ADM analog of the decomposition recalled above,
$g_{\mu\nu}^\textrm{sym}=g_{\mu\nu}^\textrm{loc PN}+h_{\mu\nu}^\textrm{nonloc}$, reads
\be
\label{eqA7}
h_{ij}^\textrm{TT sym} = h_{ij}^\textrm{TT loc PN}
+ h_{ij}^\textrm{TT nonloc}.
\ee
Here, $h_{ij}^\textrm{TT loc PN}$ is the usual,
near-zone, time-symmetric PN-expanded formally local solution
of the field equation for ${\hTT ij}$, namely
\be
\label{eqA8}
\Box{\hTT ij} = -\frac{1}{2} \delta^{\textrm{TT} kl}_{ij} J_{kl},
\ee
as taken into account in the usual ADM version of the PN expansion
\cite{Jaranowski:1997ky,Jaranowski:2015lha}.
On the other hand, the additional, tail-transported term $h_{ij}^\textrm{TT nonloc}$ is,
again, when viewed in the near zone, a \emph{homogeneous} solution
of the (linearized) field equations for ${\hTT ij}$.
The latter nonlocal, tail-transported homogeneous solution is directly deducible
from the results of Sch\"afer \cite{Schafer:1990}
who studied radiation and tail effects, within the ADM framework,
both in the wave zone and in the near zone.
The Green's functions in the latter paper [see Eq.\ (22) there]
yield the following tail-modified radiation-reaction contribution
to the (usual, retarded) solution:
\begin{multline}
\label{eqA9}
h_{ij}^\textrm{TT reac}(t) = -\frac{4G}{5c^5}
\Big[\mathcal{J}^{(1)}_{ij}(t)
\\[1ex]
+ \frac{4GM}{c^3} \int_0^{\infty}\md\tau
\ln\left(\frac{c\tau}{2s}\right)\mathcal{J}^{(3)}_{ij}(t-\tau)
\Big],
\end{multline}
where
\be
\label{eqA10}
\mathcal{J}_{ij}(t) = \int\md^3x\, J_{\langle ij\rangle}(t,\vecx)
\ee
is the trace-free part of the spatial integral of the source \eqref{eqA6}
of the field equation \eqref{eqA8} for ${\hTT ij}$.
Note that the tail-modified reactive contribution \eqref{eqA9} to ${\hTT ij}$
is simply a function of time.
This is the anti-Newtonian version of the gauge \eqref{eqA3},
namely $h_{ij}=F_{ij}(t)$ instead of $h_{00}=\frac{1}{2}F_{ij}^{(2)}x^ix^j$,
with the same linearized curvature
\begin{align*}
R^\textrm{lin}_{0i0j} = -\frac{1}{2}(\partial_{ij}h_{00}
+ \partial_{00}h_{ij} -\partial_{0i}h_{0j} - \partial_{0j}h_{0i}).
\end{align*}

The conservative, time-symmetric part of the tail-transported contribution to Eq.\ \eqref{eqA9} is
\begin{align}
\label{eqA11}
h_{ij}^\textrm{TT nonloc} &= \nonumber\\[1ex]
+\frac{8G^2M}{5c^8}
&\int_{0}^{+\infty}\md\tau
\ln\left(\frac{c|\tau|}{2s}\right) \left[ \mathcal{J}^{(3)}_{ij}(t+\tau) -\mathcal{J}^{(3)}_{ij}(t-\tau) \right]
\nonumber\\[1ex]
&= -\frac{8G^2M}{5c^8} \Pf_{2s/c}
\int_{-\infty}^{+\infty}\frac{\md\tau}{|\tau|} \mathcal{J}^{(2)}_{ij}(t+\tau).
\end{align}

The elimination of the ${\hTT ij}$ field consists
in replacing the complete solution \eqref{eqA7} of the TT field equation \eqref{eqA6}
in the Routhian \eqref{eqA5}.
At leading order, the effect of $h_{ij}^\textrm{TT nonloc}$
is linear in $h_{ij}^\textrm{TT nonloc}$
and should \textit{a priori} come from the first and third terms in Eq.\ \eqref{eqA5}.
Such a standard computation yields
\begin{multline}
\label{eqA12}
[R^\textrm{nonloc}]^\textrm{standard} = \int\md^3x \bigg[
\frac{1}{4}\partial_{k}h_{ij}^\textrm{TT nonloc}\,
\partial_{k}h_{ij}^\textrm{TT loc PN}
\\[1ex]
+ \frac{1}{4}\partial_{k}h_{ij}^\textrm{TT loc PN}\,
\partial_{k}h_{ij}^\textrm{TT nonloc}
\\[1ex]
- \frac{1}{4} h_{ij}^\textrm{TT nonloc} J_{ij} \bigg],
\end{multline}
where we exhibited the two separate (though trivially equal) contributions
coming from the (spatial) kinetic term $\frac{1}{4}(\nabla h)^2$ in Eq.\ \eqref{eqA5}.

However, the explicit expression \eqref{eqA11} of the nonlocal contribution to ${\hTT ij}$ shows,
as already mentioned, that $h_{ij}^\textrm{TT nonloc}$ is spatially independent,
being a mere function of time.
As a consequence the first two (equal) terms on the rhs of Eq.\ \eqref{eqA12} give simply a zero
contribution so that this standard computation yields
\be
\label{eqA13}
[R^\textrm{nonloc}]^\textrm{standard} = \int\md^3x \bigg[
- \frac{1}{4} h_{ij}^\textrm{TT nonloc} J_{ij} \bigg].
\ee
Actually, this answer is incorrect, because it does not agree with the one
(securely based on the equations of motion) that we used in Ref.\ \cite{Damour:2014jta}.

A formal way out of this contradiction was advocated, and used, in Ref.\ \cite{Bernard:2015njp}.
It consists in replacing the standard kinetic terms $(\partial h)^2$ in Eq.\ \eqref{eqA5}
by $-h\Box h$. If we apply this formal procedure to the spatial gradient terms in Eq.\ \eqref{eqA5}
(which are the only relevant ones at this order), i.e.,
$\frac{1}{4}(\partial_k{\hTT ij})^2\to-\frac{1}{4}{\hTT ij}\Delta{\hTT ij}$,
the result of replacing the complete solution \eqref{eqA7} in the Routhian now yields,
instead of Eq.\ \eqref{eqA12}, the new expression
\begin{multline}
\label{eqA14}
[R^\textrm{nonloc}]^\textrm{nonstandard} = \int\md^3x \bigg[
-\frac{1}{4}h_{ij}^\textrm{TT nonloc}\,\Delta h_{ij}^\textrm{TT loc PN}
\\[1ex]
- \frac{1}{4}h_{ij}^\textrm{TT loc PN}\,\Delta h_{ij}^\textrm{TT nonloc}
- \frac{1}{4} h_{ij}^\textrm{TT nonloc} J_{ij} \bigg].
\end{multline}
Formally, this is identical to Eq.\ \eqref{eqA12} modulo a (spatial) integration by parts.
However, if we use the fact that $h_{ij}^\textrm{TT nonloc}$ is simply a function of time only,
the second term on the rhs now yields a vanishing contribution,
while the first one yields (upon using the field equation for $h_{ij}^\textrm{TT loc PN}$)
\begin{multline}
\label{eqA15}
[R^\textrm{nonloc}]^\textrm{nonstandard} = \int\md^3x \bigg[
+\frac{1}{8}h_{ij}^\textrm{TT nonloc}\,\delta^{\textrm{TT} kl}_{ij} J_{kl}
\\[1ex]
- \frac{1}{4} h_{ij}^\textrm{TT nonloc} J_{ij} \bigg].
\end{multline}
If we now formally integrate by parts the spatially nonlocal projections described by
the TT projection kernel and decide that the spatial derivatives they contain yield zero
on $h_{ij}^\textrm{TT nonloc}(t)$, we further get
\begin{multline}
\label{eqA16}
[R^\textrm{nonloc}]^\textrm{nonstandard}_\textrm{formal by parts}
= \int\md^3x \bigg[
+\frac{1}{8}h_{ij}^\textrm{TT nonloc} J_{ij}
\\[1ex]
- \frac{1}{4} h_{ij}^\textrm{TT nonloc} J_{ij} \bigg]
= \int\md^3x \bigg[ -\frac{1}{8}h_{ij}^\textrm{TT nonloc} J_{ij} \bigg].
\end{multline}
As we see, we now have \emph{half} the value \eqref{eqA13}
given by using the standard kinetic terms.
And this second result now agrees with the result given in our previous work.
Indeed, the last result \eqref{eqA16} only involves the spatial integral $\mathcal{J}_{ij}$,
Eq.\ \eqref{eqA10}, of $J_{ij}$.
Inserting also the expression \eqref{eqA11} of $h_{ij}^\textrm{TT nonloc}$
as a nonlocal integral of $\mathcal{J}_{ij}^{(2)}$ yields
\begin{multline}
\label{eqA17}
[R^\textrm{nonloc}]^\textrm{nonstandard} = +\frac{G^2M}{5c^8}
\mathcal{J}_{ij}(t)
\\[1ex]
\times \Pf_{2s/c}
\int_{-\infty}^{+\infty}\frac{\md\tau}{|\tau|} \mathcal{J}^{(2)}_{ij}(t+\tau),
\end{multline}
so that the corresponding nonlocal contribution to the action reads
(modulo an integration by parts with respect to time)
\begin{multline}
\label{eqA18}
-\int R^\textrm{nonloc}_\textrm{nonstandard}\, \md t = +\frac{G^2M}{5c^8}
\\[1ex]
\times \Pf_{2s/c}
\iint \frac{\md t\,\md t'}{|t'-t|} \mathcal{J}^{(1)}_{ij}(t')\, \mathcal{J}^{(1)}_{ij}(t).
\end{multline}
Here, we note that the integral $\mathcal{J}_{ij}$ of $J_{ij}$, Eq.\ \eqref{eqA10},
is the following function of positions and momenta:
\be
\label{eqA19}
\mathcal{J}_{ij} = 2 \sum_a \frac{p_{a\langle i}p_{a j\rangle}}{m_a}
- 2 G m_1 m_2 \frac{r_{12}^{\langle i}r_{12}^{j\rangle}}{r_{12}^3},
\ee
which is simply the on-shell value of the second time derivative, say $\hat{I}^{(2)}_{ij}$
(in the notation used in Ref.\ \cite{Bernard:2015njp}),
of the quadrupole moment $I_{ij}=\sum_a m_a x_a^{\langle i}x_a^{j\rangle}$.
We thereby see that, modulo the eventual order reduction of the extra time derivative
in $\mathcal{J}^{(1)}_{ij}=\frac{\md}{\md t}\hat{I}^{(2)}_{ij}$,
the result \eqref{eqA18} agrees with the nonlocal action of Ref.\ \cite{Damour:2014jta},
as well as with the equivalent action of Ref.\ \cite{Bernard:2015njp}.
However, our explicit calculation above has given us two new results:
1) a reduced-action calculation similar to the one performed in harmonic coordinates
in \cite{Bernard:2015njp} can be done within the ADM formalism;
and 2) such a calculation has, at best, a formal value because it is very sensitive
to ill-controlled IR effects (spatial divergences) and can easily be conducted so
as to give a wrong result, as exhibited in Eqs.\ \eqref{eqA12} and \eqref{eqA13},
based on the standard $(\partial h)^2$ action.

\end{document}